\documentclass[aps,10pt,pra,twocolumn,groupedaddress,floatfix,superscriptaddress,showpacs,amsmath,amssymb]{revtex4-1}
\usepackage{amsfonts}
\usepackage{amsmath}
\usepackage{amssymb}
\usepackage{graphics,graphicx}
\graphicspath{{./}}
\usepackage{epsf}
\usepackage{subfigure}
\usepackage[%
  colorlinks=true,
  urlcolor=blue,
  linkcolor=blue,
  citecolor=blue
]{hyperref}
\usepackage[all]{hypcap}
\usepackage{color}
\usepackage[utf8]{inputenc}
\allowdisplaybreaks

\newcommand{\eps}{\varepsilon}
\newcommand{\be}{\begin{equation}}
\newcommand{\ee}{\end{equation}}
\newcommand{\bea}{\begin{eqnarray}}
\newcommand{\eea}{\end{eqnarray}}
\newcommand{\eq}[1]{Eq.~\eqref{#1}}

\newcommand{\eqss}[2]{Eqs.~\eqref{#1}-\eqref{#2}}
\newcommand{\seq}[1]{Sec.~\ref{#1}}
\newcommand{\app}[1]{App.~\ref{#1}}
\newcommand{\fig}[1]{Fig.~\ref{#1}}
\newcommand{\figs}[2]{Figs.~\ref{#1}-\ref{#2}}
\newcommand{\bem}{\begin{multline}}
\newcommand{\eem}{\end{multline}}

\newcommand{\new}[1]{#1}

\AtBeginDocument{%
    \newwrite\bibnotes
    \def\bibnotesext{Notes.bib}
    \immediate\openout\bibnotes=\jobname\bibnotesext
    \immediate\write\bibnotes{@CONTROL{REVTEX41Control}}
    \immediate\write\bibnotes{@CONTROL{%
    apsrev41Control,author="08",editor="1",pages="0",title="0",year="1"}}
     \if@filesw
     \immediate\write\@auxout{\string\citation{apsrev41Control}}%
    \fi
}%

\begin{document}

\title{Correlation-induced steady states and limit cycles in driven dissipative quantum systems}

\author{Haggai Landa}\email{haggaila@gmail.com}\affiliation{Institut de Physique Th\'eorique, Universit\'e Paris-Saclay, CEA, CNRS, 91191 Gif-sur-Yvette, France}
\affiliation{IBM Quantum, IBM Research Haifa, Haifa University Campus, Mount Carmel, Haifa 31905, Israel}
\author{Marco Schir\'o}
\email{marco.schiro@ipht.fr}\thanks{On Leave from: Institut de Physique Th\'{e}orique, Universit\'{e} Paris Saclay, CNRS, CEA, F-91191 Gif-sur-Yvette, France}
\affiliation{JEIP, USR 3573 CNRS, Coll\`ege de France,   PSL  Research  University, 11,  place  Marcelin  Berthelot,75231 Paris Cedex 05, France}
\author{Grégoire Misguich}\email{gregoire.misguich@cea.fr}
\affiliation{Institut de Physique Théorique, Université Paris-Saclay, CEA, CNRS, 91191 Gif-sur-Yvette, France}
\affiliation{Laboratoire de Physique Théorique et Modélisation, CNRS UMR 8089, Université de Cergy-Pontoise, 95302 Cergy-Pontoise, France}

\begin{abstract}
We study a driven-dissipative model of spins one-half (qubits) on a lattice with nearest-neighbor interactions. Focusing on the role of spatially extended spin-spin correlations in determining the phases of the system, we characterize the spatial structure of the correlations in the steady state, as well as their temporal dynamics. In dimension one we use essentially exact matrix-product-operator simulations on large systems, and pushing these calculations to dimension two, we obtain accurate results on small cylinders. We also employ an approximation scheme based on solving the dynamics of the mean field dressed by the feedback of quantum fluctuations at leading order. This approach allows us to study the effect of correlations in large lattices with over one hundred thousand spins, as the spatial dimension is increased up to five. In dimension two and higher we find two new states that are stabilized by quantum correlations and do not exist in the mean-field limit of the model. One of these is a steady state with mean magnetization values that lie between the two bistable mean-field values, and whose correlation functions have properties reminiscent of both. The correlation length of the new phase diverges at a critical point, beyond which we find emerging a new limit cycle state with the magnetization and correlators oscillating periodically in time.

\end{abstract}

\maketitle

\section{Introduction}\label{Sec:Intro}

Atomic, optical, and solid-state systems are often operated in many-body nonequilibrium regimes characterized by a competition between interactions, nonlinearity, coherent external driving and dissipative dynamics. These include arrays of coupled circuit quantum electrodynamic (QED) units \cite{AndrewNatPhys}, spin ensembles embedded into large optical or
microwave cavities \cite{KuboetalPRL10,GrezesEtAlPRX14,FinkEtAlPRX17}, mesoscopic quantum circuits of increasing complexity interfaced with microwave resonators \cite{BruhatEtAlPRX16,ParlavecchioEtAlPRL15}, trapped ions \cite{bohnet2016quantum} and cold atoms \cite{Tomadin_prl10, keeling_collective_2010}.
A rich pattern of behaviors at the interface between quantum optics and condensed matter physics is observed with systems of strong light-matter interactions \cite{ChangEtAlNatPhys06,bhaseen_dynamics_2012,
OtterbachEtAl_PRL13,HoningEtAlPRA13,CarusottoCiutiRMP13,RMP_Esslinger13,
RMP_Esslinger13,SchmidtKochAnnPhy13,LeHurReview16,
NohAngelakisRepProgPhys2016,HartmannJOpt2016,SchiroPRL16}. 
Dissipative phase transitions and critical phenomena in open systems attract increasing attention and research activity  \cite{Greentree_NatPhys2006,Hartmann_NatPhys2006, nagy_critical_2011,oztop_excitations_2012,
schiro_phase_2012,schiro_quantum_2013, torre_keldysh_2013,kulkarni_cavity-mediated_2013,brennecke_real-time_2013,SchiroPRL16,FitzpatrickEtAlPRX17,
ScarlatellaFazioSchiroPRB19,MaEtAlNature19,
MarinoDiehlPRB16,shchadilova2018fermionic, doi:10.1002/qute.201800043,PhysRevA.100.042113,munoz2019spontaneous}.
 Driving and dissipation can be utilized in the generation of topological quantum states by coupling to a specially tailored bath \cite{bardyn2013topology} or time-periodic (Floquet) driving \cite{annurevOkaKitamura}. 
Proposals to realize artificial gauge fields with circuit-QED photonic lattices \cite{Koch_PRA11}, chiral edge modes \cite{Petrescu_PRA12,HafeziFanMigdallTaylor2013} and quantum Hall fluids of light \cite{Angelakis_FQH_PRL08,umucalilar_artificial_2011, HafeziLukinTaylor_NJP2013}, follow the paradigm of engineering topological states in fully neutral quantum systems \cite{rechtsman2013photonic}.

An open quantum system coupled to a Markovian bath obeys a Lindblad master equation for the time evolution of the density matrix $\rho$,
\be
\partial_t\rho = -i[H,\rho]+\mathcal{D}[\rho],\qquad \hbar=1.\label{Eq:dtrho}
\ee
Here the first term on the right-hand side describes the coherent evolution due to interactions and possibly coherent driving terms (with the Hamiltonian $H$ in the rotating frame), while the 
dissipator $\mathcal{D}[\rho]$ accounts for dephasing and relaxation processes due to the environment, described by a set of jump operators. 
The simplest approach for obtaining solutions of such systems, applied to various driven-dissipative systems (mostly of coupled spins or oscillators) \cite{lee_antiferromagnetic_2011,qian_phase_2012, PhysRevLett.110.163605,jin_steady-state_2014,SchiroPRL16,PhysRevA.97.053616,foss-feig_emergent_2017,biondi_nonequilibrium_2017, chan_limit-cycle_2015,wilson_collective_2016, marcuzzi2014universal}, is the mean-field (MF) decoupling limit, in which $\rho$ is approximated by a product of single-site density matrices.
 The MF phase diagrams obtained this way manifest both translationally-invariant steady states and antiferromagnetic (AF) or staggered phases of a spontaneously broken symmetry, where neighbouring sites have different mean magnetization or density. Oscillatory limit cycle (LC) phases that break time-translation invariance~\cite{IeminiEtAlPRL18} have also been found, and in addition, there are bistable or multistable parameter regions where two or more different states coexist.

A highly debated question in the literature is whether the AF, LC, and multistable phases found in MF are genuine features of the quantum system.

A large number of studies concern one-dimensional (1D) lattices (mostly with nearest-neighbor interactions), finding that the MF AF phase is replaced by a uniform phase stabilized by AF correlations
\cite{lee_antiferromagnetic_2011, wilson_collective_2016,biondi2017spatial,biondi_quantum_2018}, and bistability is replaced by a smooth crossover accompanied by large quantum fluctuations \cite{weimer_variational_2015,PhysRevA.93.023821,foss-feig_emergent_2017,vicentini_critical_2018,bimodality}. These conclusions rely on accurate numerical methods that can be applied to large 1D systems, such as matrix product operator (MPO) simulations, but an experimental investigation is still lacking.
In two-dimensional (2D) lattices, numerical methods are much more limited. MF bistability has been found by some approximate methods to be replaced by a sharp first order jump between two phases \cite{foss-feig_emergent_2017,vicentini_critical_2018, weimer_variational_2015,JinEtAlPRB18, KshetrimayumEtalNatComm2017}. A dynamical timescale diverging at the jump has been found \cite{weimer_variational_2015,vicentini_critical_2018, bimodality}, which is attributed to a vanishing Liouvillian gap -- the smallest magnitude of the real part of the nonzero eigenvalues \cite{PhysRevA.98.042118}. Limit cycles in the driven-dissipative Heisenberg lattice which are predicted in MF, were found to disappear due to short-range correlations in finite dimensions \cite{owen_quantum_2018}.

In \cite{bimodality} we have presented a theoretical scenario for quantum bistability in driven dissipative lattice systems of spatial dimension two and higher. We have provided numerical evidence in its support using a self-consistent theory of quantum fluctuations beyond mean field, dubbed MF with quantum fluctuations (MFQF) and MPO simulations, applied to spins one-half. Within this scenario the MF bistability is not washed away by quantum correlations, provided the thermodynamic limit of infinite large system size is taken before the long-time limit, i.e. provided the system is studied on time scales which are smaller than exponential in system size. We also discussed what this scenario implies concerning the slowly-relaxing eigenstates of the Liouvillian super operator.

The questions we address in this work are:
(i) How the patterns of stationary states and fixed point of the dissipative dynamics change with increasing interactions in the system, and in particular how they deviate from the MF solutions,
(ii) whether phases not accessible in MF can be induced and stabilized in the system by the quantum correlations, and (iii) whether long-range (spatial and temporal) order induced by a  competition between driving and dissipation can be stabilized by quantum fluctuations beyond MF.
Specifically, we consider a driven-dissipative quantum spin model, discussed in \cite{bimodality}. We study its steady state and dynamical properties for larger values of the nearest-neighbor interactions (as compared to the regime studied in \cite{bimodality}), presenting for completeness and as a reference point, a detailed study of the MF limit.
In dimension two and higher we find two new states that are stabilized by quantum correlations and do not exist in the MF limit of the model. One of these is a steady state with mean magnetization values that lie between the two bistable MF values, and whose correlation functions have properties reminiscent of both. The correlation length of the new phase diverges at a critical point, beyond which we find emerging a new limit cycle state with the magnetization and correlators oscillating periodically in time.

The paper is organized as follows.  In \seq{Sec:MainResults} we present a summary of the main results of the paper.
In \seq{Sec:Model} we present the equations of motion which form the starting point for calculating the dynamics of observables within the MF and MFQF approaches, presenting details of the MF limit
in \seq{Sec:MF}. 
In \seq{Sec:MFQF} we discuss the MFQF approximation scheme that goes beyond MF, and in \seq{Sec:MPO} we describe a method based on matrix product operators (MPO), applied to the present model in 1D and 2D. In \seq{Sec:Numerics} we present our results obtained using MPO and MFQF, and in \seq{Sec:Outlook} we conclude with a summary and outlook. The Appendix contains some further details of the theory and numerics.

\section{Main Results}\label{Sec:MainResults}

We study a driven-dissipative model on a hypercubic lattice in $D$ dimensions, with $N$ sites located at $R\in \mathbb{Z}^D$, and connectivity $\mathcal{Z}=2D$. 
We consider spins one-half using the Pauli matrices at each site, $\sigma_R^a$ with $a=\{x,y,z\}$, and the ladder operators $\sigma^{\pm}_R = ({\sigma^{x}_R\pm i\sigma^y_R})/{2}$.
Decomposing the Hamiltonian into the kinetic (hopping) part $T$ and the sum of on-site terms, we have
\be H = T +\sum_{R}\left(\frac{\Delta}{2}\sigma_R^z+ \Omega\sigma_R^x\right),\label{Eq:HR}\ee
where $H$ describes two-level quantum systems driven with amplitude $\Omega$ and detuning $\Delta$, in a frame rotating with the drive, using the rotating wave approximation, as derived in \app{App:Rotating}. Here, $T$ is 
\bem T = -\sum_{\langle R,R'\rangle} \left(J\sigma^+_R \sigma^-_{R'} +{\rm H.c.} +\frac{1}{2} J_z \sigma^z_R \sigma^z_{R'}\right)=\\ -\frac{1}{2}\sum_{\langle R,R'\rangle} \left(J\sigma^x_R \sigma^x_{R'} +J\sigma^y_R \sigma^y_{R'} +J_z \sigma^z_R \sigma^z_{R'}\right),\label{Eq:Tdef}\end{multline}
with the summation extending over all lattice bonds.
Setting $J=0$ we have a dissipative Ising model, introduced in \cite{lee_antiferromagnetic_2011}.
In this summary section we set $J_z=0$, focusing on the driven-dissipative XY model, first considered in \cite{PhysRevA.93.023821,wilson_collective_2016}.
For independent couplings with rate $\Gamma$ of each site to a zero-temperature bath, the dissipator in \eq{Eq:dtrho} reads 
\be
 \mathcal{D}[\rho]={\Gamma}\sum_R\left( \sigma_R^-\rho \sigma_R^{+}-\frac{1}{2}\left\{\sigma_R^{+}\sigma_R^-,\rho\right\}\right).\ee

\begin{figure}
{\includegraphics[clip,width=0.48\textwidth]{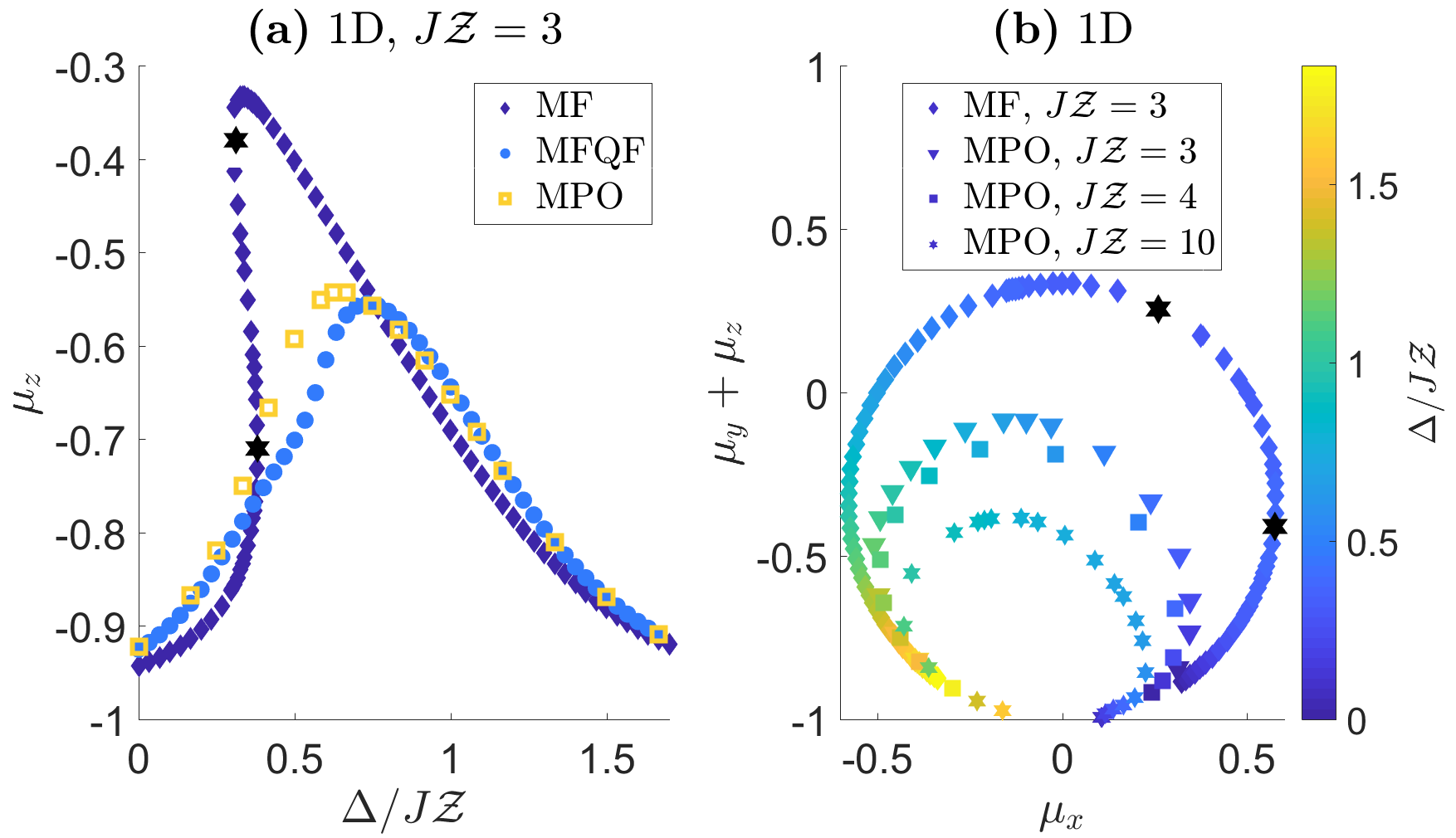}}
\caption{(a) Mean steady-state $z$ magnetization $ \mu_z^S$, as a function of $\Delta/J\mathcal{Z}$ for $\Gamma=1$, $\Omega=0.5$ and $J\mathcal{Z}=3$, on a 1D chain. The MF limit manifests bistability, with three coexisting solutions, two of which (those on the branches coming from the limits $\mu_z\to -1$ as $\Delta\to\{0,\infty\}$), are stable. Two black stars mark the points where the unstable branch meets each of the two stable ones. An accurate treatment using MPO shows a crossover occurring within a range of detuning shifted from the MF bistability region. This crossover is also approximately captured by MFQF, incorporating quantum fluctuations at leading order, and dressing MF. (b) The steady-state magnetization in the plane defined by $\mu_z^S=\mu_y^S-1$, with the color code denoting $\Delta/J\mathcal{Z}$, for the MF and MPO at a few values of $J\mathcal{Z}$. The deviation from the ellipse forming the locus of the MF solutions results directly from correlations.
} \label{Fig:Comparison}
\end{figure}

Assuming a translationally invariant state (further discussed in \seq{Sec:Model}), we define the time-dependent mean magnetization and its steady-state value (assuming it exists) as 
\be \mu_a(t)\equiv \left\langle \frac{1}{N} \sum_R \sigma_R^a\right\rangle = \left\langle\sigma_R^a\right\rangle,\quad \vec\mu^S(t)\equiv \lim_{t\to\infty}\vec\mu(t)\label{Eq:mu}.\ee
As shown in \seq{Sec:MFsteady}, the steady-state magnetization lies in the plane $\mu_z^{S}=2 \mu_y^S\Omega /\Gamma-1$ that for $\Gamma=1$ and $\Omega=0.5$ becomes $\mu_z^S=\mu_y^S-1$, which can be spanned by the two (orthogonal) vectors $\hat\mu_x$ and $\hat\mu_y+ \hat\mu_z$. This is an exact result that requires only translation-invariance.

In MF, $\vec\mu^S$ is further constrained to lie on the circumference of an ellipse. Figure \ref{Fig:Comparison} depicts the MF trajectory of $\vec\mu^S$ in the steady state plane for $J\mathcal{Z}=3$ as a function of $\Delta/J\mathcal{Z}$ (with this rescaling facilitating comparison at different interaction strengths discussed in the following).
As seen in \fig{Fig:Comparison}(a), for $\Delta\approx 0$ there is a single solution with $\mu_z^S\approx -1$ and $\mu_x^S>0$. Increasing $\Delta$, this solution moves counter-clockwise along the MF ellipse [\fig{Fig:Comparison}(b)]. At $\Delta/J\mathcal{Z}\approx 0.3$, a new stable solution appears at a high $\mu_z^S$ value, together with an unstable solution. As $\Delta$ is increased, the new stable solution moves counter-clockwise along the MF ellipse while the unstable solution moves clockwise, until at $\Delta/J\mathcal{Z}\approx 0.37$, the unstable solution collides with the first stable solution, both becoming complex and hence ceasing to be physical solutions. The remaining single solution continues along the ellipse towards $\mu_z^S\to -1$ as $\Delta\to\infty$.

In the presence of correlations, the magnetization departs from the MF ellipse (but remains in the plane). 
MPO calculations in 1D with up to a few hundred sites (and the results verified for convergence with $N$) show a significant deviation from  MF for $0.2\lesssim\Delta/J\mathcal{Z} \lesssim 1.4$, with
a smooth crossover between the two limiting regimes. This crossover can be seen in \fig{Fig:Comparison}(a) [for $J\mathcal{Z}=3$], showing also that the MFQF approximation is capable of washing away the bistability region resulting in a single phase that follows approximately the numerically exact MPO solution. This result has been discussed in \cite{bimodality} for a somewhat higher value of $J\mathcal{Z}=4$ (and identical values of the other parameters). We attribute this capability of MFQF to
the fact that it incorporates correlations  with a nontrivial spatial dependence, which is an important characteristic of the many-body solution.
As shown in \cite{bimodality}, in the heart of the crossover region the correlations grow by up to a few orders of magnitude. The strength of the correlations depends naturally on the interaction coefficient $J$, and in \fig{Fig:Comparison}(b) it can be seen that, as $J$ is increased, the trajectory of $\vec\mu^S$ deviates further from the MF ellipse, due to the correlators increasing in magnitude.
In \seq{Sec:Numerics} we present a detailed comparison of the MFQF and MPO solutions in 1D for a larger $J$ value ($J\mathcal{Z}=10$), and discuss the similarities and deviations observed. 

\begin{figure}
{\includegraphics[clip,width=0.48\textwidth]{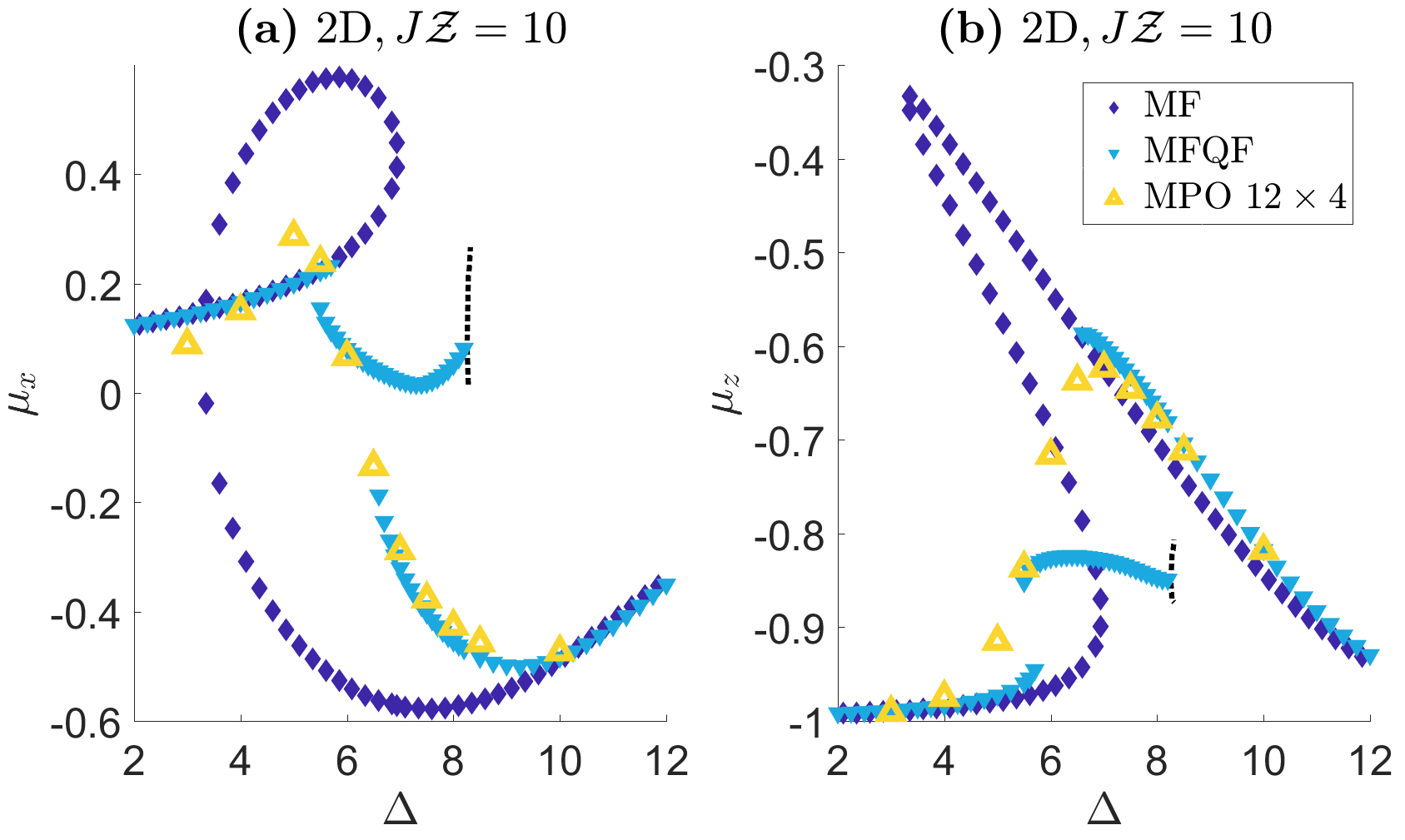}}
\caption{(a) $ \mu_x^S$, and (b) $ \mu_z^S$, as a function of $\Delta$, for fixed values of the other paramaters, $\Gamma=1$, $\Omega=0.5$, $J=2.5$, on a 2D lattice ($J\mathcal{Z}=10$). MF manifests bistability for $3.4\lesssim\Delta \lesssim 7$. The MFQF approximation predicts multistability with a new branch of an emergent phase appearing, whose magnetization values are intermediate between the MF branches. Beyond the right edge of this branch appears a stable limit cycle in a small region $8.24\lesssim \Delta\lesssim 8.32$, marked by a black dotted line, giving the amplitude of oscillations (which vary with $\Delta$). See the text for a detailed discussion, and \fig{Fig:Correlations2D} for an analysis of the MFQF correlation functions.} \label{Fig:Comparison052D}
\end{figure}

\begin{figure}
\includegraphics[width=\linewidth]{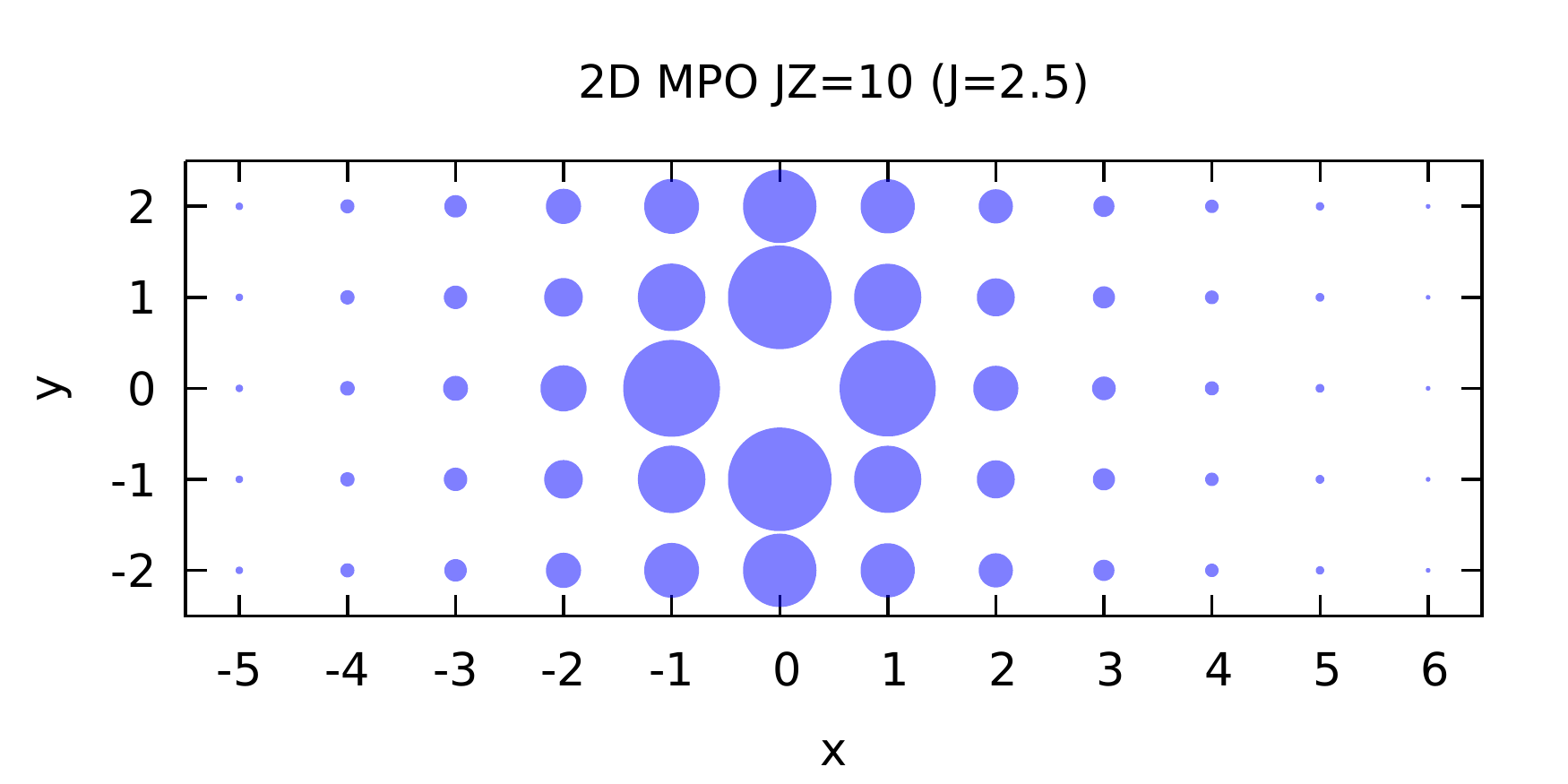}
\caption{The connected two-point correlation function $\eta_{xx}(R)$ [see \eq{Eq:etadef}] in the steady state obtained by MPO simulations on a 2D system (4$\times$12 cylinder), for the same parameters as in \fig{Fig:Comparison052D},
and $\Delta=7$.
The circle diameter on the site $R=\{x,y\}$ is proportional to the absolute value of the correlation between the site at the origin with that in $R$.
The MPO parameters and the approach to simulate 2D systems are discussed in \seq{Sec:MPO} and depicted in \fig{Fig:eta_mpo}.
}
\label{Fig:eta_mpo0}
\end{figure}

Turning to 2D lattices, \fig{Fig:Comparison052D} shows the magnetization of the MF and MFQF steady states for strong interactions (${J}\mathcal{Z}=10$, and $\Gamma=1$, $\Omega=0.5$) on a 2D lattice with periodic boundary conditions, and the MPO steady state on a cylinder of length $12$ and circumference $4$. The MPO mean magnetization coincides quantitatively very well with that of MFQF (simulating a 2D lattice of up to $200^2$ sites), for the steady state on the branch coming from high $\Delta$ down to $\Delta\approx 7$. This is a regime where the correlation length is not large (of order $1-2$ lattice sites). The steady state on the cylinder appears to be locally very close to that of a large system. This is evidenced by the correlation functions, and \fig{Fig:eta_mpo0} depicts the connected two-point correlation function $\eta_{xx}(R)$ [defined in \eq{Eq:etadef}], calculated in MPO. We see that the nearest-neighbour correlations $\eta_{xx}(|R|=1)$ are nearly isotropic, and along the cylinder's symmetry axis the 
correlations decay rapidly.
However, for $5.5 \lesssim \Delta\lesssim 8.2$ MFQF predicts bistability in the thermodynamic limit, while due to the finite size of the MPO cylinder, the MPO steady state is necessarily unique, and must depend smoothly on $\Delta$.
We can therefore not expect these finite-size MPO calculations to show bistability or even a discontinuity. 
At low $\Delta$ values incommensurate spin-spin correlations develop (with a long correlation length, although with a relatively small magnitude). There, the MPO simulations are affected by the small size  of the cylinder, and the agreement with MFQF is only semi-quantitative.

\begin{figure}
{\includegraphics[clip,width=0.48\textwidth]{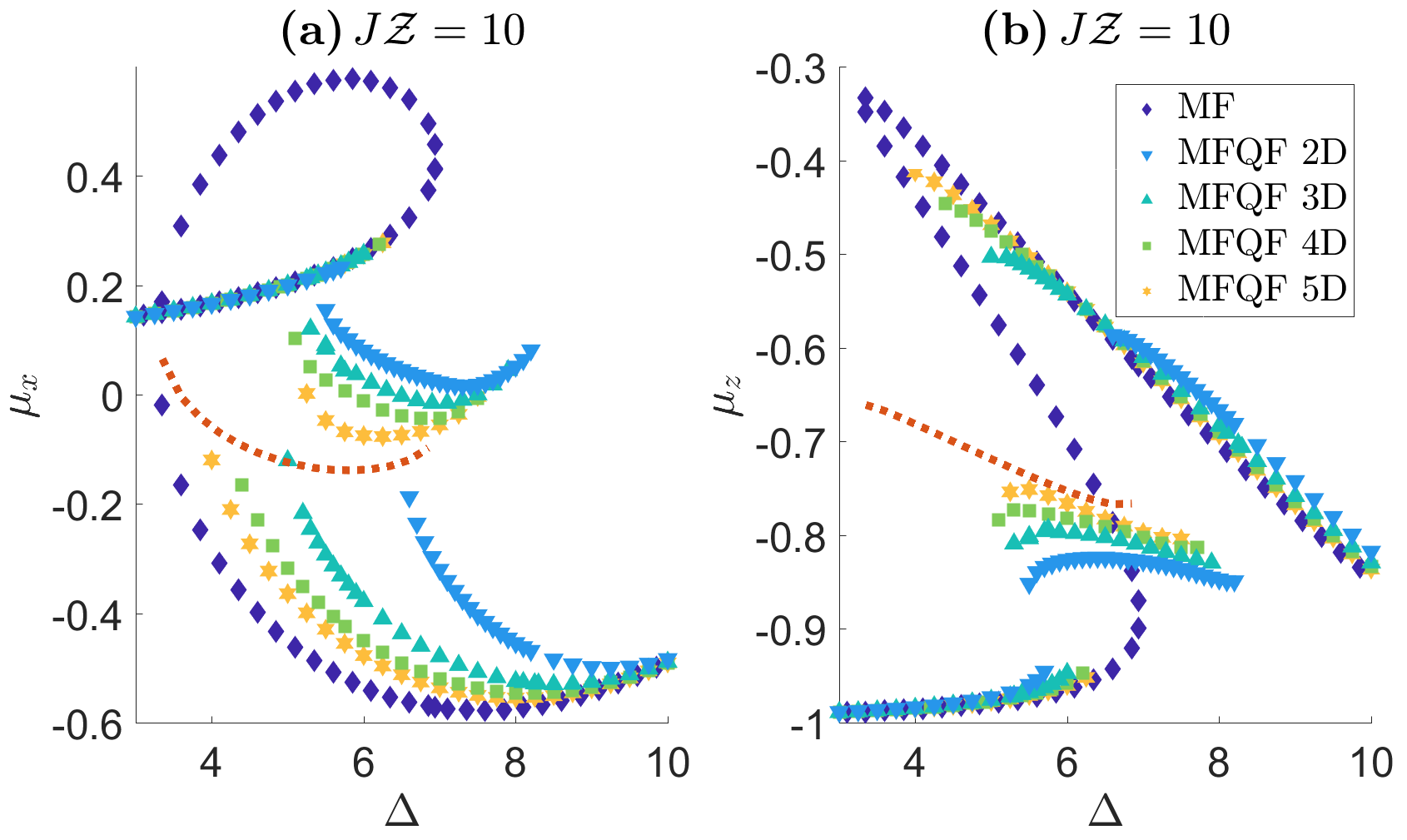}}
\caption{(a) $ \mu_x^S$, and (b) $ \mu_z^S$, as a function of $\Delta$, for MFQF in 2D-5D, and MF. The parameters are as in \fig{Fig:Comparison052D}, keeping $J\mathcal{Z}$ fixed by varying $J$ with the dimension. As the dimension increases, the two MF-like branches progressively converge towards the MF branches in a larger parameter region, while the intermediate branch gradually shifts and shrinks. Beyond the right edge of this branch for 2D-4D appears a stable limit cycle (not indicated), in a small regime of $\Delta$ that also shrinks with the dimension. The dotted red line denotes the arithmetic average of the two bistable MF phases. The simulations were run with lattices of up to $200^2$, $40^3$, $20^4$, and $10^5$ sites for 2D-5D respectively (and periodic boundary conditions).} \label{Fig:Comparison5D}
\end{figure}

\begin{figure}[t]
\subfigure[\,2D]{
\includegraphics[trim = 0cm 0cm 0.cm 0cm, clip,width=0.23\textwidth]{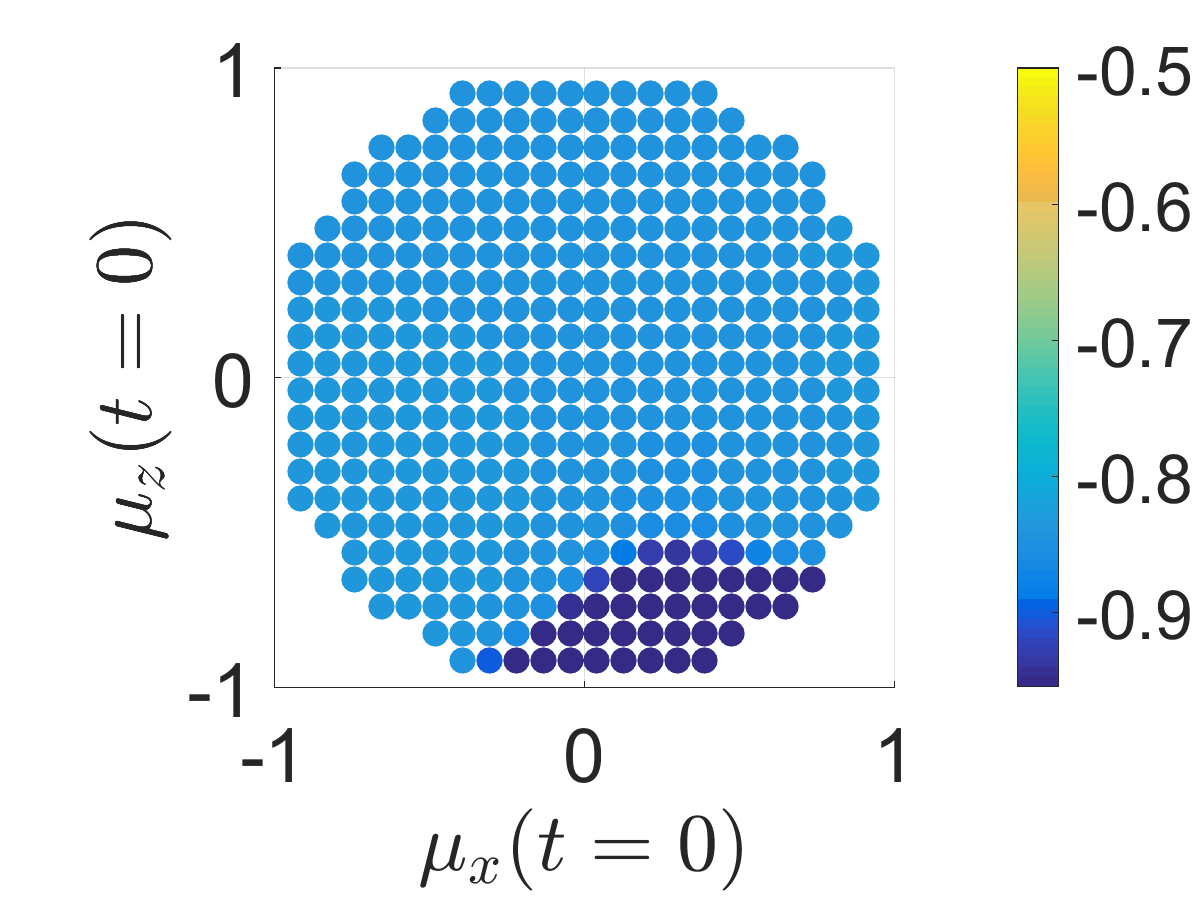}}
\subfigure[\,3D]{\includegraphics[trim = 0cm 0cm 0.cm 0cm, clip,width=0.23\textwidth]{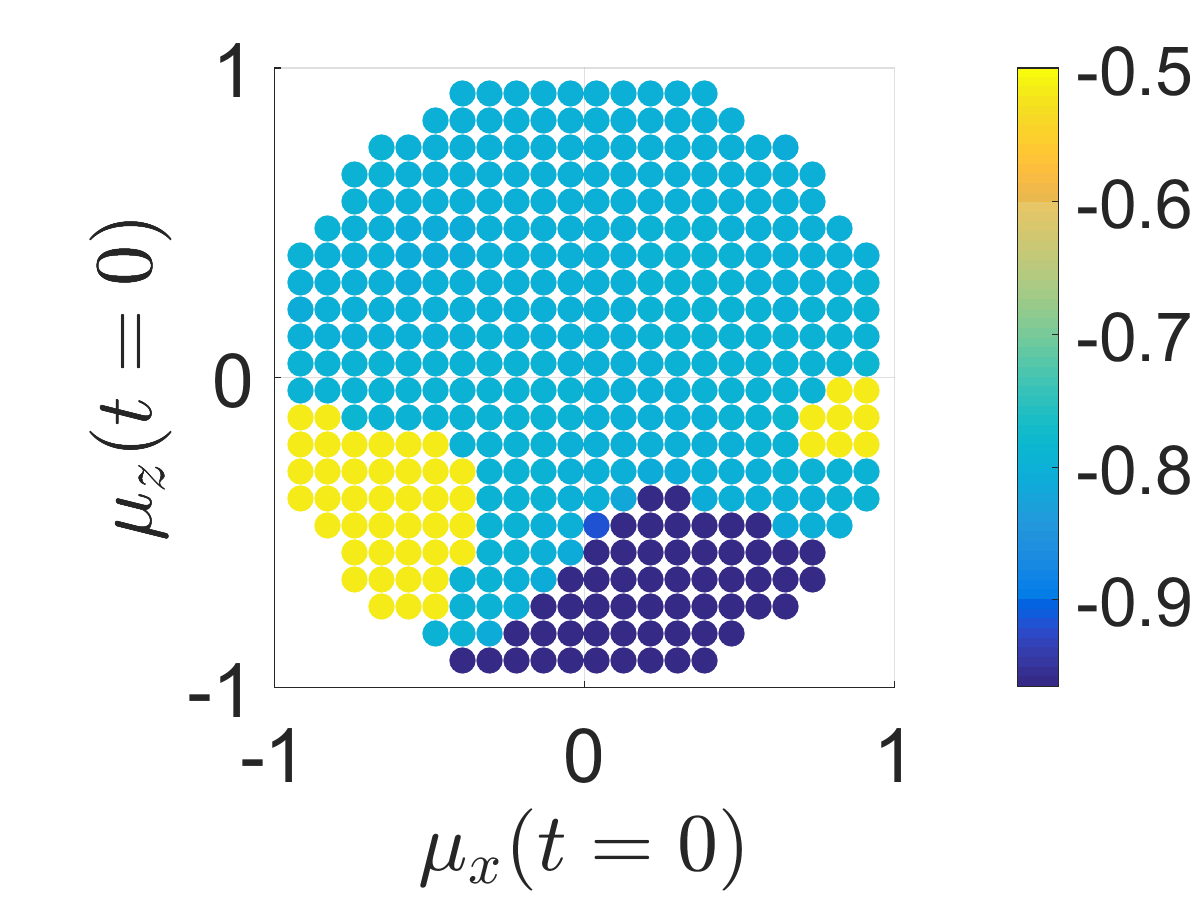}}
\bigskip
\subfigure[\,4D]{\includegraphics[trim = 0cm 0cm 0.cm 0cm, clip,width=0.23\textwidth]{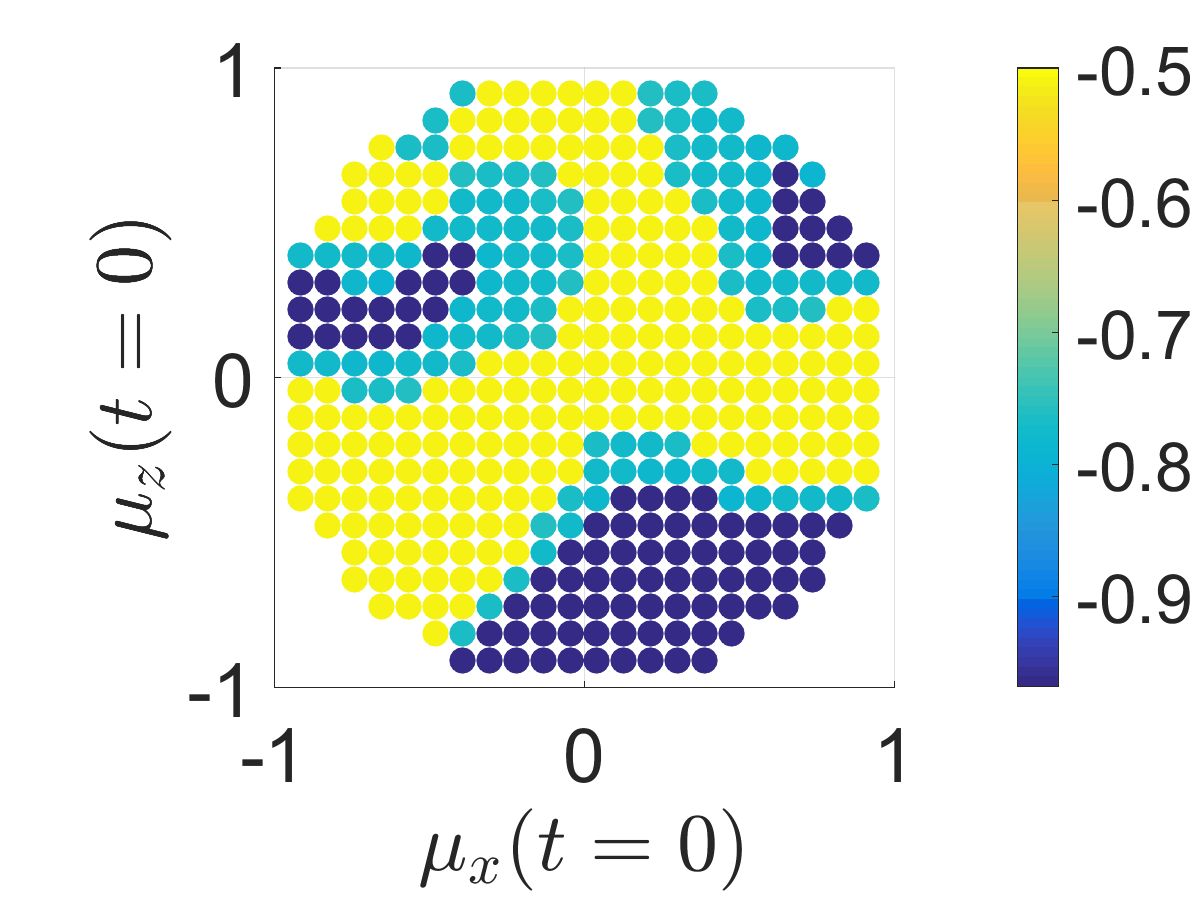}}
\subfigure[\,MF]{\includegraphics[trim = 0cm 0cm 0.cm 0cm, clip,width=0.23\textwidth]{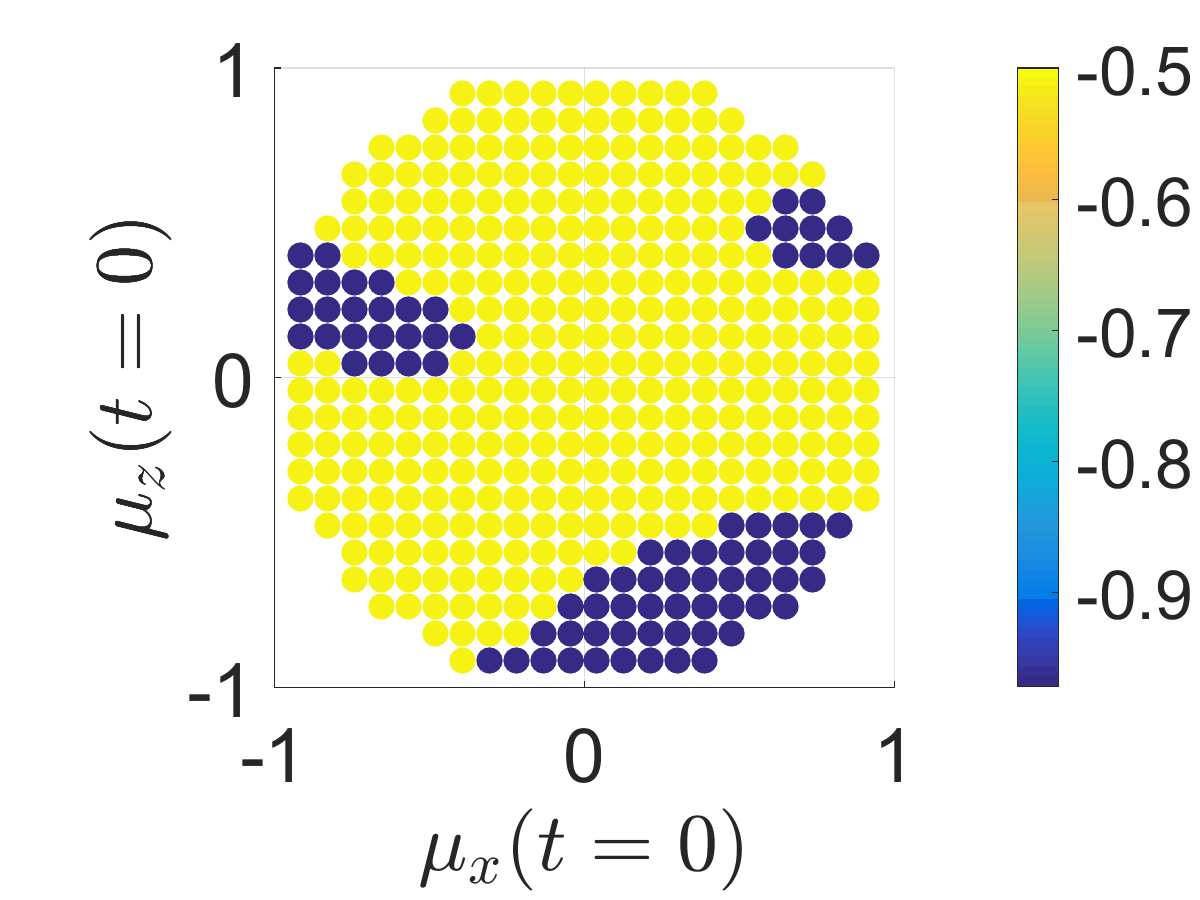}}
\caption{Basins of attraction of the multistable steady states, depicted by $\mu_z^S$ in the final state (given by the color code), as a function of different initial conditions $\vec \mu(t=0)$ started within a transversal cut through the unit-magnetization sphere (see \seq{Sec:MFDynamics} for details), for $\Gamma=1$, $\Omega=0.5$, $J\mathcal{Z}=10$, and $\Delta=5.6$. As the dimension is increased for successive panels with (a) 2D, $60^2$ spins, (b) 3D, $20^3$ spins, (c) 4D, $10^4$ spins -- all simulated using MFQF -- the basin of attraction of the middle branch (light blue hues) progressively shrinks and the plots plausibly converge towards the MF basins, shown in panel (d) with two bistable states.} \label{Fig:BasinsD}
\end{figure}

In addition to the possibility of the coexistence of stable MF-like branches, in MFQF a new branch appears in some range of $\Delta$, as shown in \fig{Fig:Comparison052D}. The mean magnetization on this branch has intermediate values between the two MF-like branches. The new intermediate branch has large and spatially modulated two-point correlations, whose characteristics are shown in detail in \fig{Fig:Correlations2D} in \seq{Sec:Numerics}. Moreover, we find that at the high-$\Delta$ edge of the branch the correlation length diverges (in practice, reaches the linear size of the lattice). Beyond this point, we find a small range of $\Delta$ values for which an oscillatory LC state is stabilized by large correlations extending throughout the lattice. This phenomenon is not present in the MF approximation. The oscillation patterns of the mean magnetization and two-point correlation functions are presented in \seq{Sec:Numerics}.

The MFQF approximation is easy to apply in  higher dimensions, and the dynamics can be solved with a very large number of spins. Figure \ref{Fig:Comparison5D} shows the results of such simulations carried out with large lattices from 2D through 5D. We find multistable branches in progressively larger ranges of $\Delta$, which converge towards the MF bistability region. 
At the same time, the $\Delta$ range of the intermediate branch is slowly shrinking as $D$ is increased.
We thus see how the MFQF solutions gradually approach the MF ones when $D$ is increased.
For the steady state we present in \fig{Fig:BasinsD} a look into the basins of attraction of the coexisting multistable steady-states. The basins of attraction, whose construction starting from initial product states is explained in more detail in \seq{Sec:MFDynamics}, are taken at fixed values of the parameters, varying only $D$. We see that, as $D$ is increased from  2 to 4, the basin of attraction of the new branch shrinks and gives way to increasingly MF-like basins for the two MF states.

Our study proposes some answers to the questions posed in the outset. Briefly, our study suggests that
quantum phases in presence of driving and dissipation can support large fluctuations, depending on the dimension and interaction strength, and that a behaviour commonly associated with classical nonlinearity -- namely bistability and hysteresis -- is effectively possible in the thermodynamic limit of the studied quantum system in 2D and above.
We find new emerging states of the system in the long time limit, phases not accessible in MF, which are induced and stabilized by quantum fluctuations and correlations. For critical parameters long-range spatial order can be sustained in the lattice due to the competition of the drive, dissipation and interactions, and for some parameter ranges, also a temporal order in the form of a spontaneous forming of a stable limit cycle. Whether these phases survive as true solutions of the full quantum system remains a fundamental open question, possibly awaiting for experimental quantum simulation for full confirmation. Experiments are foreseeable with trapped ions and superconducting qubits, and can possibly answer general questions about the dynamics of many-body quantum systems, beyond the sizes accessible to state-of-the-art numerics.

\section{Equations of Motion}\label{Sec:Model}

\new{ In this section we present the equations of motion (e.o.m) for observables of the quantum system \cite{SandriSchiroFabrizioPRB12}, which form the basis for the MF and MFQF approaches. We define $n$-points expectation values in the form 
\be\langle \sigma_{R_1}^{a}\sigma_{R_2}^{b}\cdots \sigma_{R_n}^{c}\rangle\equiv {\rm Tr}\{\rho\sigma_{R_1}^{a}\sigma_{R_2}^{b}\cdots \sigma_{R_n}^{c}\}.\ee 
By multiplying \eq{Eq:dtrho} with an operator $O$ and taking the trace, we get an e.o.m for the expectation value
\be \partial_t{\rm Tr}\{\rho O\} = -i{\rm Tr}\{[H,\rho]O\}+{\rm Tr}\{\mathcal{D}[\rho]O\}\label{Eq:eomTr}.\ee
Starting with single-site operators $O$, this leads to a hierarchy of equations that depend on the value of correlators at the next order, $n+1$. 
A simple way to handle the derivation is to use the linearity of the equation and treat separately the Hamiltonian and the dissipative parts. Matrix elements do not depend on the picture by which they are calculated, and in the following we calculate the Hamiltonian part of the e.o.m in the Heisenberg picture, and the dissipative part in the Schr\"{o}dinger picture.}

 An Heisenberg e.o.m for any operator $O$ reads in the absence of dissipation,
\be \left.\partial_t O\right|_{\Gamma=0} = i[H,O],\ee
and using the commutation relations of \app{App:eom} we obtain the Heisenberg e.o.m (for $\Gamma=0$),
\bea &\left.\partial_t \sigma_R^x\right|_{\Gamma=0} =& -\sum_{\langle R'\rangle}\left(J \sigma_{R'}^y \sigma_R^z - J_z \sigma_{R'}^z \sigma_{R}^y \right)  -{\Delta} \sigma_R^y ,\label{Eq:eomsigmaRx} \\ 
 &\left.\partial_t \sigma_R^y\right|_{\Gamma=0} =& \sum_{\langle R'\rangle} \left(J \sigma_{R'}^x \sigma_R^z -J_z \sigma_{R'}^z \sigma_{R}^x \right)-2\Omega\sigma_R^z +{\Delta} \sigma_R^x,\nonumber \label{Eq:eomsigmaRy} \\ 
& \left.\partial_t \sigma_R^z\right|_{\Gamma=0} =& -J\sum_{\langle R'\rangle} \left(\sigma_{R'}^x \sigma_R^y - \sigma_{R'}^y \sigma_{R}^x\right) +2 \Omega \sigma_R^y \label{Eq:eomsigmaRz},
\eea
with the summation of ${\langle R'\rangle}$ extending over the nearest neighbours of the lattice site $R$. 

We now assume that the initial density matrix commutes with spatial translations and reflection. In this case, with a Hamiltonian that is also invariant under these operations, the time evolution will remain in the same symmetry sector, which is characterized by a uniform magnetization, and two-point correlations that are only a function of the distances. This precludes the possibility of spontaneous symmetry breaking in the thermodynamic limit, however for $\Delta>0$ no AF phase is expected based on MF and 1D MPO results. In addition, we did not observe any sign of modulation instability of the uniform states.  
We (re-)define the MF magnetization of \eq{Eq:mu},
\be \mu_a(t)\equiv \left\langle \frac{1}{N} \sum_R \sigma_R^a\right\rangle = \left\langle\sigma_R^a\right\rangle,\ee
and we define a two-point correlation function (correlator),
\be \vartheta_{ab}(R,R',t)\equiv \left\langle \sigma_{R}^a\sigma_{R'}^b\right\rangle,\qquad R\neq R',\ee
which is a function of the difference $R-R'$ alone, symmetric in $a,b$ (because $\sigma_{R'}^a$ and $\sigma_R^b$ commute). The connected  two-point correlator is defined (for $R\neq R'$) by
\bem \eta_{ab}(R,R',t)\equiv\\ \left\langle \left(\sigma_{R}^a -\mu_a\right)\left(\sigma_{R'}^b-\mu_b\right)\right\rangle=\vartheta_{ab}(R,R',t) - \mu_a\mu_b,\label{Eq:etadef}\end{multline}
We will similarly refer to the connected three-point correlator defined for $R\neq R'\neq R''$,
\be \zeta_{abc}(R,R',R'',t) \equiv \left\langle \left(\sigma_{R}^a -\mu_a\right)\left(\sigma_{R'}^b-\mu_b\right)\left(\sigma_{R''}^c-\mu_c\right)\right\rangle,\label{Eq:tilderhoR0}
\ee
which is again a function of the differences only.

Substituting the definition of the correlators in \eqss{Eq:eomsigmaRx}{Eq:eomsigmaRz}, taking the expectation value and including the dissipative terms obtained by calculating ${\rm Tr}\{\mathcal{D}[\rho]\sigma_R^a\}$ in the Schr\"{o}dinger picture as in \eq{Eq:eomTr}, we get the e.o.m
\bea \partial_t\mu_x &=& -\left({J}-J_z\right)\mathcal{Z}\vartheta_{yz}(1) -{\Delta} \mu_y -\frac{\Gamma}{2}\mu_x,\label{Eq:eomRmag_x} \\
\partial_t\mu_y &=& \left({J}-J_z\right)\mathcal{Z} \vartheta_{xz}(1) -2\Omega\mu_z +{\Delta} \mu_x -\frac{\Gamma}{2}\mu_y, \label{Eq:eomRmag1} \\ 
\partial_t \mu_z &=& 2 \Omega \mu_y - {\Gamma} (1 +\mu_z),\, \label{Eq:eomRmag2}
\eea
where ${\vartheta}_{ab}(1)$ is the correlator at distance $\|R-R'\|_1=1$ (with $\|v\|=\sum_{j=1}^D|v_j|$ the $l_1$ norm on the $D$-dimensional cubic lattice). We write explicitly 
\be {\vartheta}_{ab}(1)= \mu_a \mu_b +\eta_{ab}(1).\ee
\eqss{Eq:eomRmag_x}{Eq:eomRmag2} are exact, but do not form a closed system. In the following sections we study the MF and MFQF approximations, obtained by closing the equations by truncation at different orders of $n$-point correlations. For a related approach based on expansion of the density matrix in the inverse connectivity, see \cite{biondi2017spatial}.

\section{Mean field}\label{Sec:MF}

\subsection{The mean-field steady state}\label{Sec:MFsteady}

Setting $\eta\to 0$ in \eqss{Eq:eomRmag_x}{Eq:eomRmag2}, which amounts to assuming that the density matrix is a product of identical on-site states, we get the MF e.o.m,
\bea \partial_t\mu_x = -{J}\mathcal{Z}\mu_y \mu_z -{\Delta} \mu_y -\frac{\Gamma}{2}\mu_x,\label{Eq:dotmux_mf} \\
\partial_t\mu_y = {J}\mathcal{Z} \mu_x \mu_z -2\Omega\mu_z +{\Delta} \mu_x -\frac{\Gamma}{2}\mu_y,\label{Eq:dotmuy_mf} \\ 
\partial_t\mu_z = 2 \Omega  \mu_y-{\Gamma} (1 +\mu_z),\label{Eq:dotmuz_mf}
\eea
where we have set $J_z=0$; Since the MF equations depend only on the difference $J-J_z$, the MF results for the XY model hold equally well for the Ising model (by the replacement $J\to -J_z$), and equivalently, for any combination of the two interaction types (this equivalence ceases to hold when correlations beyond MF are considered). Combining \eqss{Eq:dotmux_mf}{Eq:dotmuz_mf}, the squared length of the MF spin evolves according to
\be \partial_t (\vec{\mu}^2) =-\Gamma\left(\mu_x^2+\mu_y^2+2\mu_z^2 +2\mu_z\right)\label{Eq:dotmu2},\ee
where the r.h.s is always negative for magnetization within the unit sphere.
The MF equations are invariant under two different discrete symmetries
\be J \to -J,\qquad \Delta \to -\Delta,\qquad \mu_x \to -\mu_x,\label{Eq:sym1}\ee
and
\be  \Omega \to - \Omega,\qquad \mu_x \to - \mu_x,\qquad \mu_y \to -\mu_y.\label{Eq:sym2}\ee
These two symmetries manifest themselves in the parameter-space dependence of the steady state as discussed below.

Setting the time-derivative of \eqss{Eq:dotmux_mf}{Eq:dotmuz_mf} to zero and isolating $\mu_y$, its steady-state value is determined through a third order polynomial equation, which can have either one real root (and two complex conjugate roots), or three real ones.
Each root of the  polynomial gives a solution for $\mu_y$, from which we get immediately the corresponding
$\mu_x$ and $\mu_z$.
In the case of three different real roots, there are two stable solutions and one unstable solution, as we find from a linear stability analysis; 
linearizing the e.o.m about any steady-state solution by substituting
\be \vec\mu = \vec{\mu}^S +\delta\vec\mu,\ee
we obtain a linear system for the small fluctuations which is defined by the matrix
\be \left(\begin{array}{ccc} -\Gamma/2 &-J\mathcal{Z} {\mu}_z^S-\Delta & -J \mathcal{Z}{\mu}_y^S \\ J \mathcal{Z}{\mu}_z^S +\Delta &-\Gamma/2 &J \mathcal{Z}{\mu}_x^S -2\Omega\\ 0 & 2\Omega &-\Gamma \end{array}\right),\ee
and $\vec{\mu}^S$ is linearly stable to uniform perturbations when all eigenvalues of this matrix have a nonpositive real part.

\begin{figure} {\includegraphics[clip,width=0.43\textwidth]{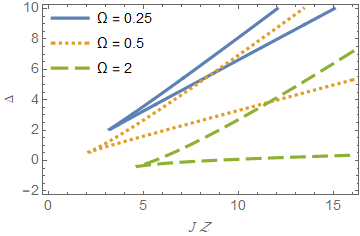}} \caption{MF bistability regions in $(J\mathcal{Z},\Delta)$ parameter plane, for $\Gamma=1$  and three values of $\Omega$. Within the regions bounded by the two lines at a fixed $\Omega$, two stable steady-state solutions coexist together with an unstable one. The unstable solution coincides with the one stable solution at region boundaries (on the curves shown), and the end point of the two curves forms a critical point.} \label{Fig:Bistability} \end{figure}

We note that the limit of $\Gamma\to 0$ is singular, because for $\Gamma=0$ there is no steady-state and the dynamics become Hamiltonian. We present all results by setting $\Gamma=1$, which defines the units of energy and time.  
 The condition for bistability is that the discriminant of the cubic equation of the magnetization is positive. As a function of the parameters, the discriminant is a high order polynomial, and the condition of its positivity defines a 3D region in $(\Omega,J,\Delta)$ parameter space, at  fixed value of $\Gamma$.  This region has disconnected components related by the symmetries of \eqss{Eq:sym1}{Eq:sym2}. We find that bistability requires $|J\mathcal{Z}|> 2$ (see below). For any fixed $\Omega$ there is a bistability region in the $(J,\Delta)$-plane on each side of the line $J\mathcal{Z}=-\Delta$, starting at a cusp point from which two curves emanate defining the bistability boundary, see \fig{Fig:Bistability}. On each (bifurcation) curve the unstable solution coincides with one stable solution, and both solutions lead to a zero eigenvalue upon linearization, with the same eigenvector. At the cusp the three MF solutions coincide and each has a zero eigenvalue, all with the same eigenvector.
The cusp is also a critical point where an effective $\mathbb{Z}_2$ symmetry appears, as discussed in \cite{marcuzzi2014universal}.

\begin{figure} {\includegraphics[clip,width=0.45\textwidth]{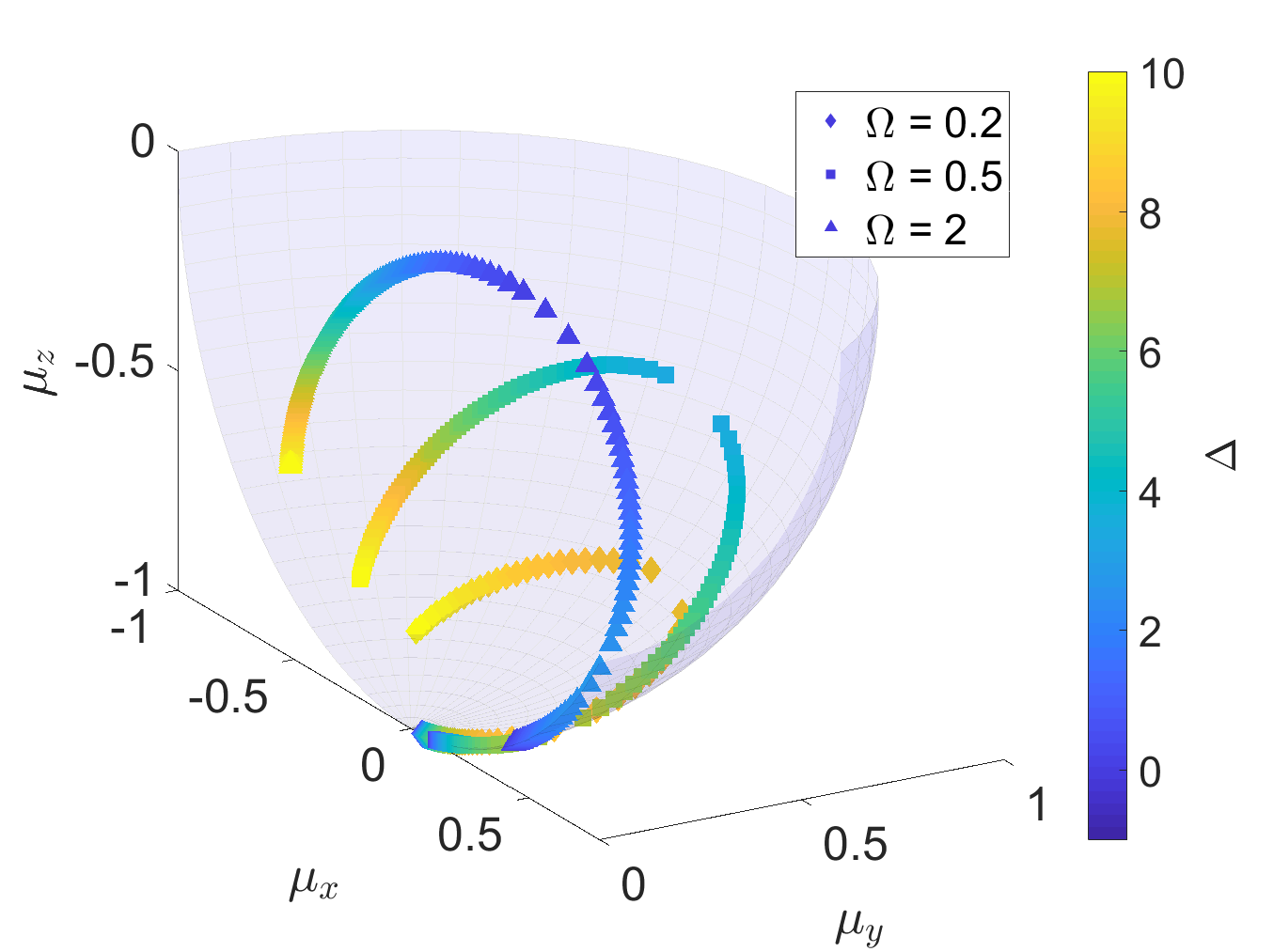}} \caption{MF steady-state trajectories in $\vec\mu$-space as a function of $\Delta$ (given by the color code), for $\Gamma=1$, $J\mathcal{Z}=10$, and three values of $\Omega$. At any fixed $\Omega$, the steady-state solutions obtained at all $J\mathcal{Z}$ and $\Delta$ values form an ellipse within a plane whose inclination is determined by $\Omega$, with the sign of $\mu_y$ equal to that of $\Omega$. See the text for a detailed discussion.} \label{Fig:Traj} \end{figure}

However, not all values of $\vec{\mu}$ are allowed in the steady state. Using \eq{Eq:dotmu2}, the mean magnetization vector in the steady state is constrained to lie on the surface of the ellipsoid
\be  \left(\frac{\mu_x^S}{1/\sqrt{2}}\right)^2+ \left(\frac{\mu_y^S}{1/\sqrt{2}}\right)^2+\left(\frac{\mu_z^S+1/2}{{1/2}}\right)^2=1,\label{Eq:ellipsoid}\ee
which is centered about $\mu_z=-1/2$ and has a $z$ principal semi-axis of length $1/2$, going from the infinite temperature solution point at $\vec\mu= 0$ to the pure (product) state at $\mu_z=-1$, so that at the steady state the solution obeys $-1\le \mu_z^S\le 0$. 
Using \eq{Eq:dotmuz_mf}, the steady state at a fixed value of $\Omega$ is obtained by the intersection of the ellipsoid of \eq{Eq:ellipsoid} with the plane
\be \mu_z^S=2 \mu_y^S\Omega /\Gamma-1,\label{Eq:mu_z_linear}\ee
which also shows that $\mu_z^S$ is related to $\mu_y^S$ by a displacement and stretching, and that the sign of $\mu_y$ is equal to the sign of $\Omega$. 
For $\Omega=0$ the steady state is the pure product state $\mu_z^S=-1$, while for for $|\Omega|\to\infty$ (with the other parameters fixed), we have found that the steady state is the infinite temperature state, $\vec\mu^S\to \vec 0$. 

\begin{figure}
{\includegraphics[clip,width=0.45\textwidth]{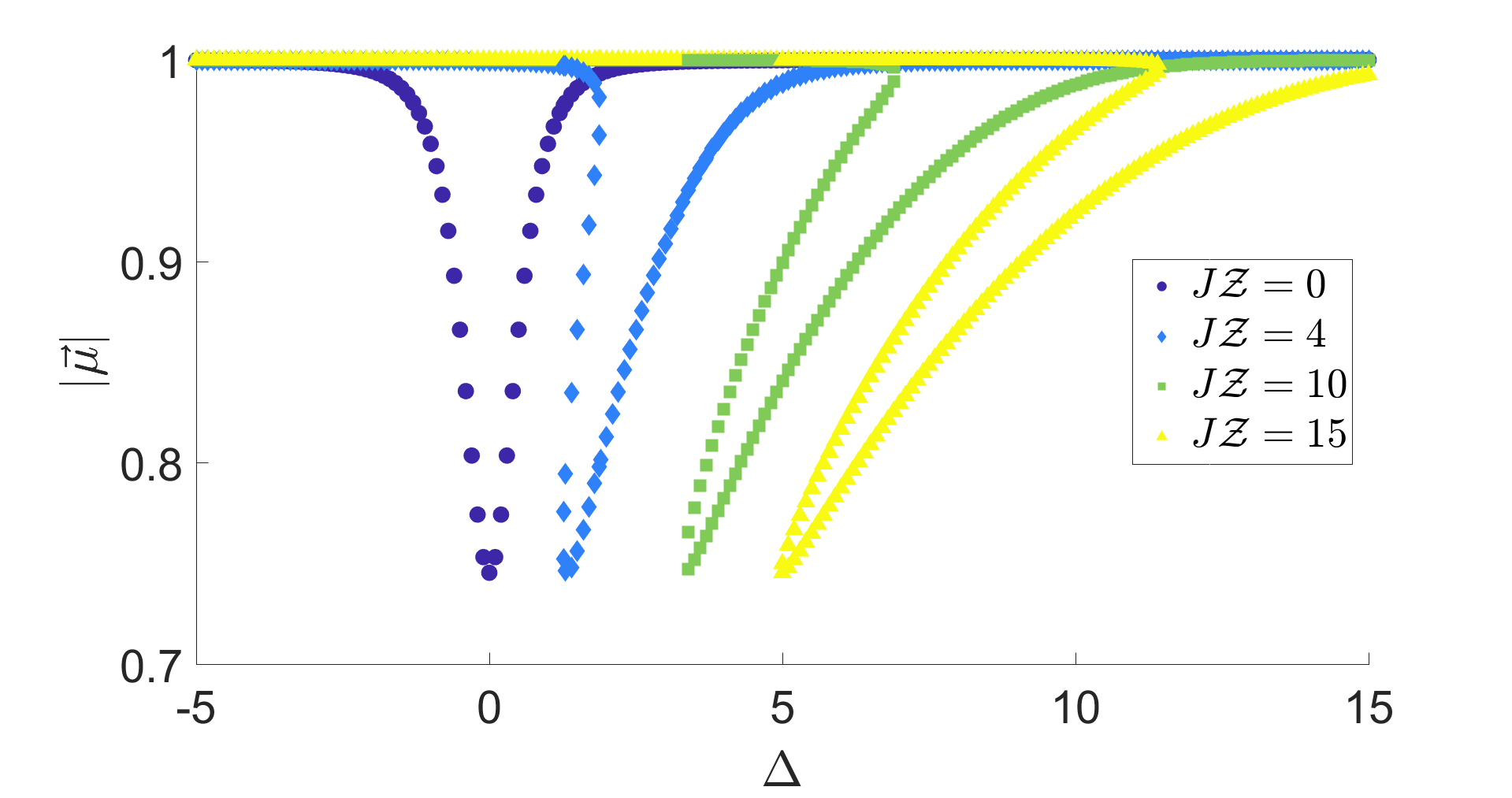}}
\caption{$|\vec{\mu}^S|$ as a function of $\Delta$ for $\Gamma=1$, $\Omega = 0.5$, and a few values of $J\mathcal{Z}$. The bistability range in $\Delta$ increases and shifts to higher values when increasing $J\mathcal{Z}$. Due to the constraints of \eqss{Eq:ellipsoid}{Eq:mu_z_linear}, the minimal norm $|\vec{\mu}^S|$ of the magnetization (which can be considered as a distance from the infinite-temperature state $|\vec{\mu}^S|=0$), is bounded.
} \label{Fig:absmuOmega05}
\end{figure}

Figure \ref{Fig:Traj} shows how the three components of the mean-field magnetization vary with $\Delta$ for fixed $J\mathcal{Z}=10$ and three values (0.2, 0.5 and 2) of $\Omega$. The ellipsoid of \eq{Eq:ellipsoid} is visible, as well as the fact that the magnetization vector must lie in a plane determined by \eq{Eq:mu_z_linear}. Pairs of points (on the same ellipsoid) with the same color ({\it i.e.} same $\Delta$) are the bistable solutions. However, as this is hard to discern in this figure, \figs{Fig:absmuOmega05}{Fig:absmuDelta1} present $|\vec{\mu}^S|$ as a function of some of the model parameters, exemplifying the limits of $\mu^S$, the bistability, and its dependence on the parameters. Some further properties of the MF steady state are derived in \app{App:MFApp}.

\begin{figure}
{\includegraphics[clip,width=0.45\textwidth]{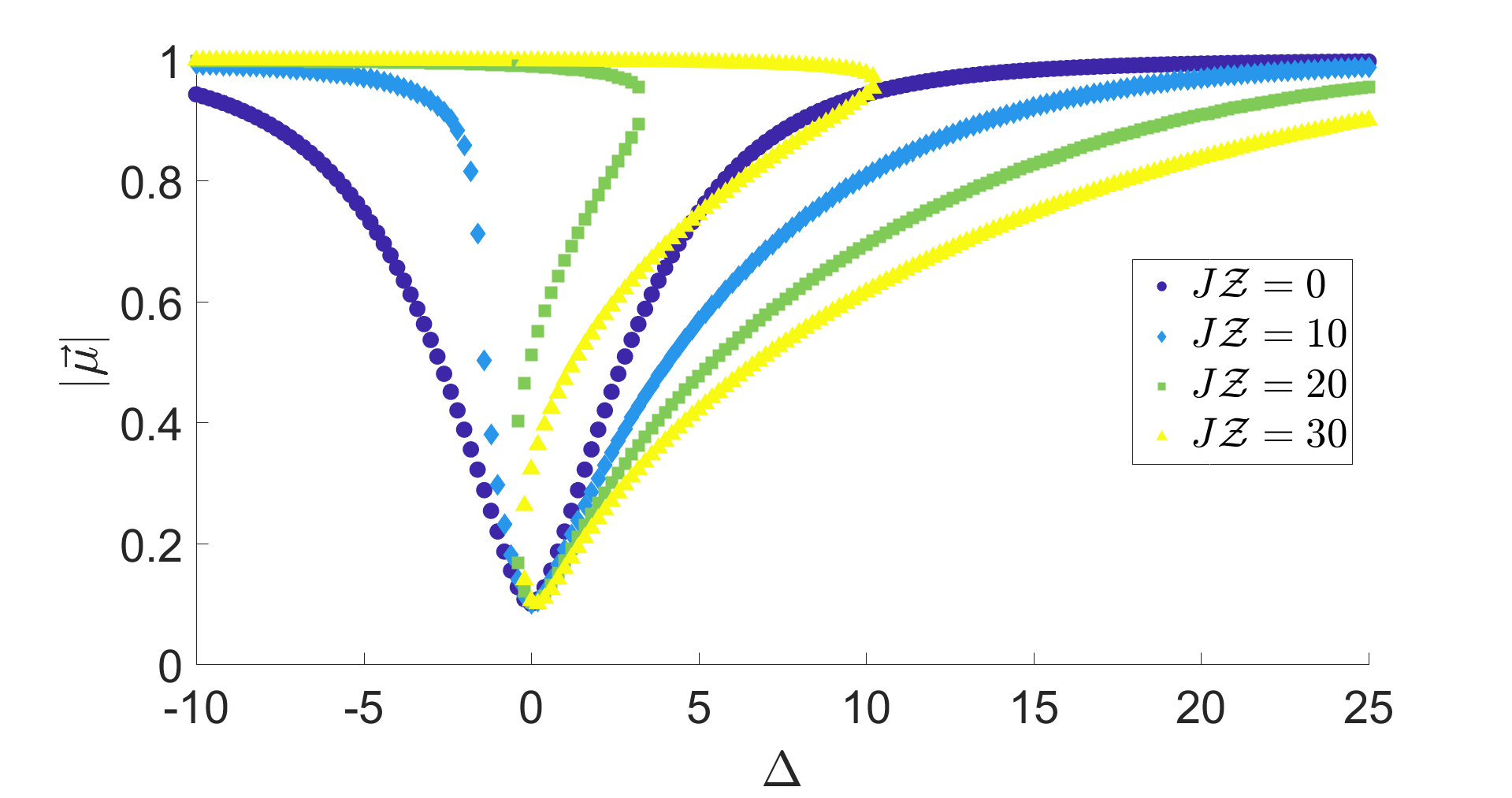}}
\caption{As in \fig{Fig:absmuOmega05}, for $\Omega = 5$. For these parameters, $\vec{\mu}^S$ approaches the infinite-temperature state around $\Delta=0$.
} \label{Fig:absmuOmega5}

\end{figure}

\begin{figure}[t!]
{\includegraphics[clip,width=0.45\textwidth]{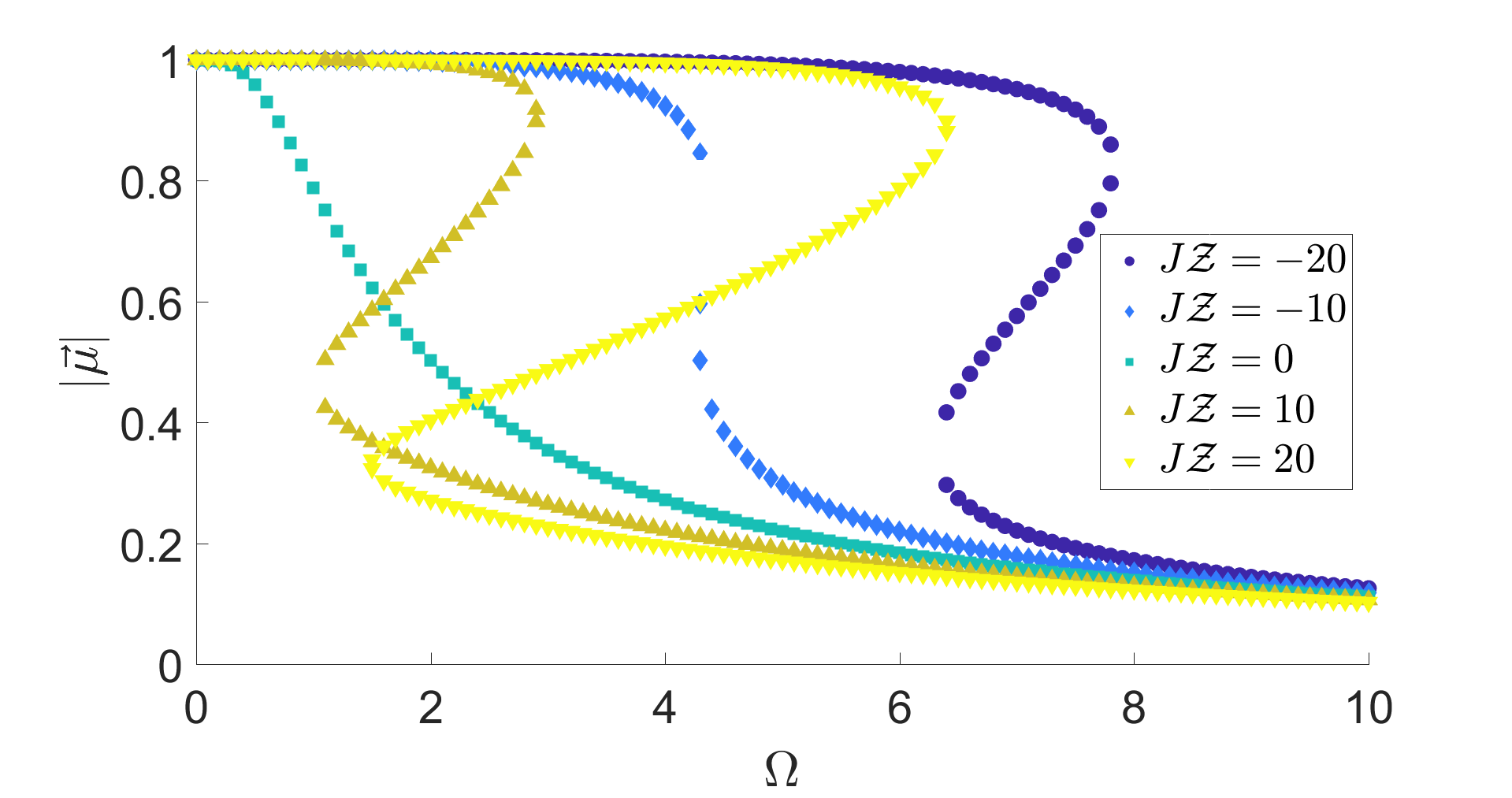}}
\caption{$|\vec{\mu}^S|$ as a function of $\Omega$ for $\Gamma=1$, $\Delta = 1$, and a few values of $J\mathcal{Z}$. The figure is symmetric for $\Omega\to -\Omega$, and it can be seen how at fixed values of the other parameters, $|\vec\mu|\to 0$ as $|\Omega|\to\infty$.} \label{Fig:absmuDelta1}
\end{figure}

\subsection{Mean-field dynamics}\label{Sec:MFDynamics}

We now turn to the dynamics associated to the MF e.o.m. We here focus on  one property of the dynamical system, which is the distribution of initial conditions converging to the possible steady states for parameters in the bistability region.
The basins of attraction of each of the bistable solutions can be calculated by starting the dynamics at initial conditions chosen within the unit magnetization sphere $\vec\mu^2=1$, and following the dynamics to the steady state. The basins of attraction contain information on the global dynamics and were presented in \fig{Fig:BasinsD} of \seq{Sec:MainResults} using the MFQF approach.

\begin{figure}
\subfigure[\,$\Omega=0.5,J\mathcal{Z}=4,\Delta=1.6$]{
\includegraphics[trim = 0cm 0cm 0.cm 0cm, clip,width=0.23\textwidth]{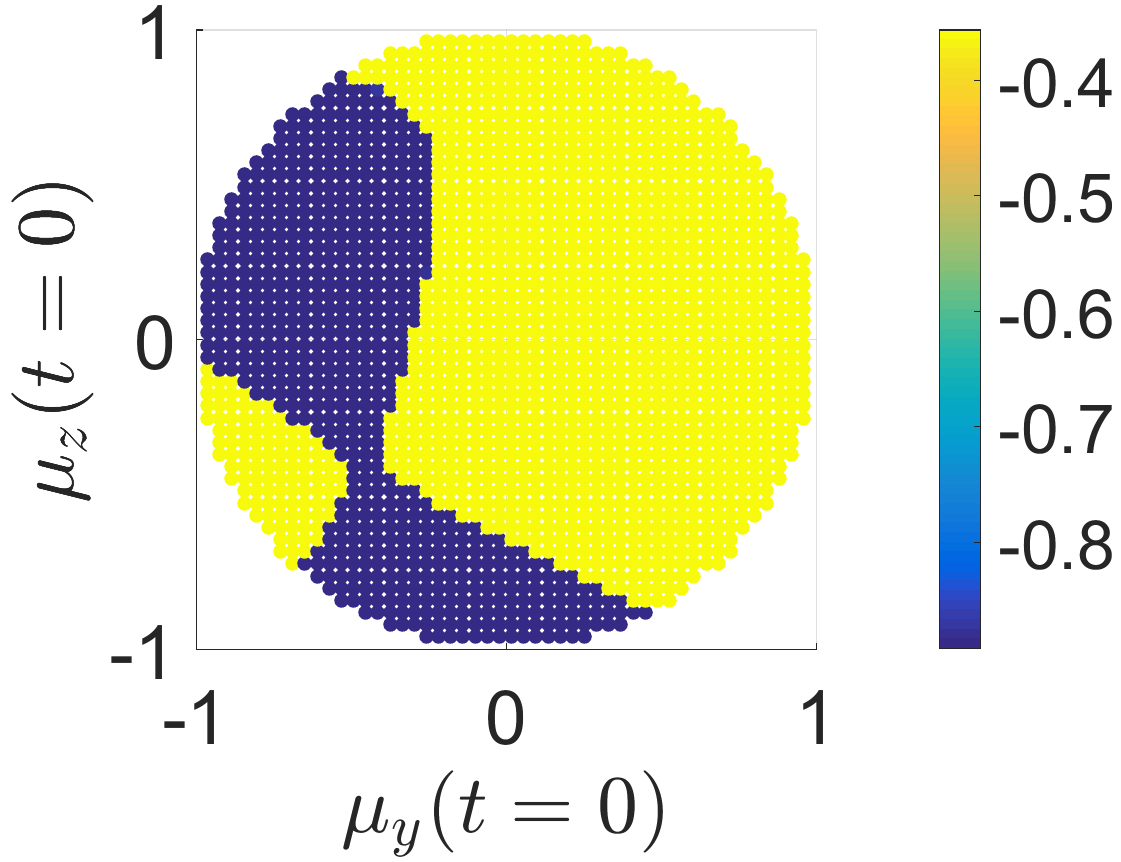}}
\subfigure[\,$\Omega=0.5,J\mathcal{Z}=16,\Delta=8$]{\includegraphics[trim = 0cm 0cm 0.cm 0cm, clip,width=0.23\textwidth]{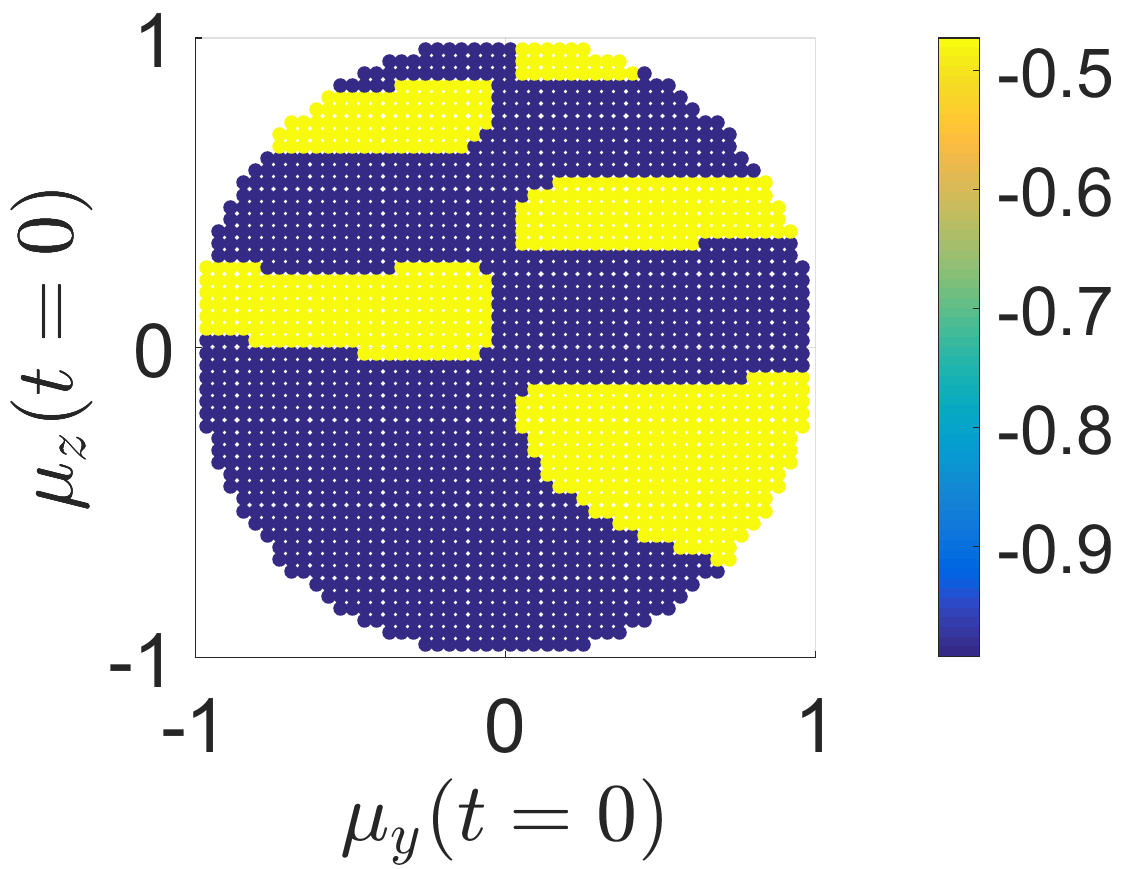}}
\bigskip
\subfigure[\,$\Omega=2,J\mathcal{Z}=50,\Delta=14$]{\includegraphics[trim = 0cm 0cm 0.cm 0cm, clip,width=0.23\textwidth]{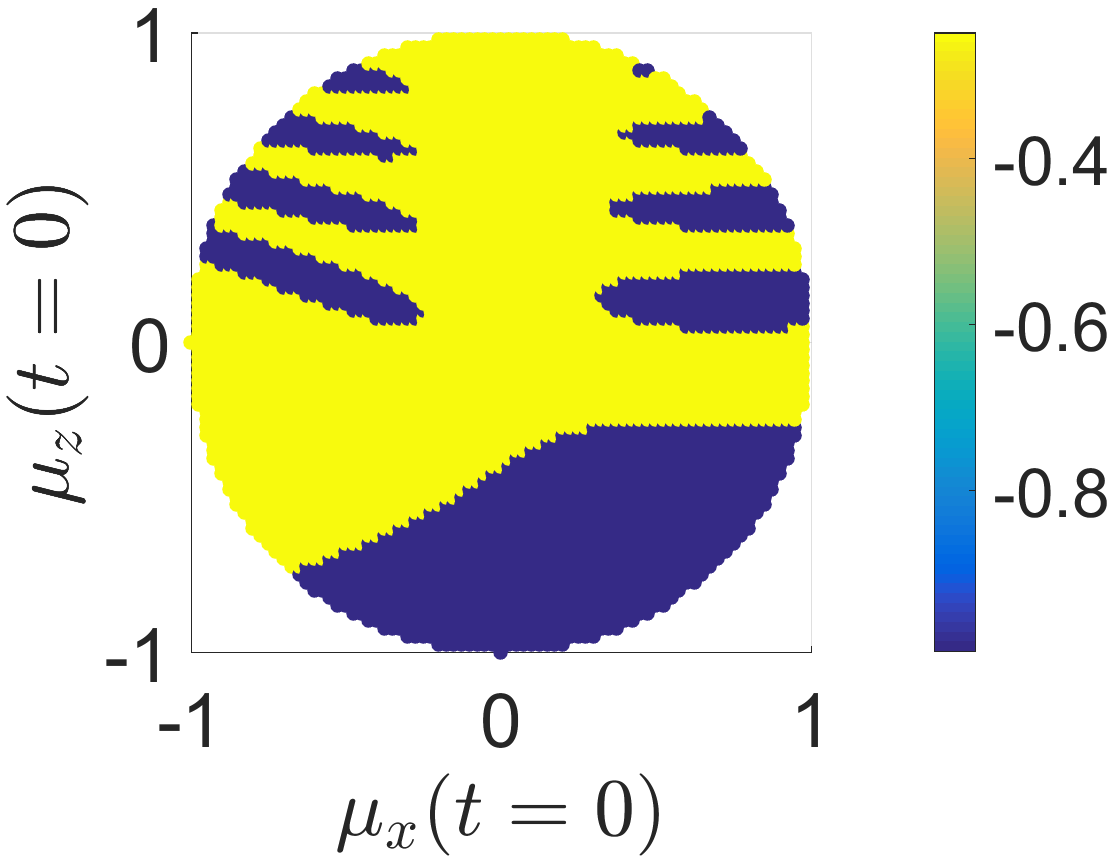}}
\subfigure[\,$\Omega=10,J\mathcal{Z}=220,\Delta=20$]{\includegraphics[trim = 0cm 0cm 0.cm 0cm, clip,width=0.23\textwidth]{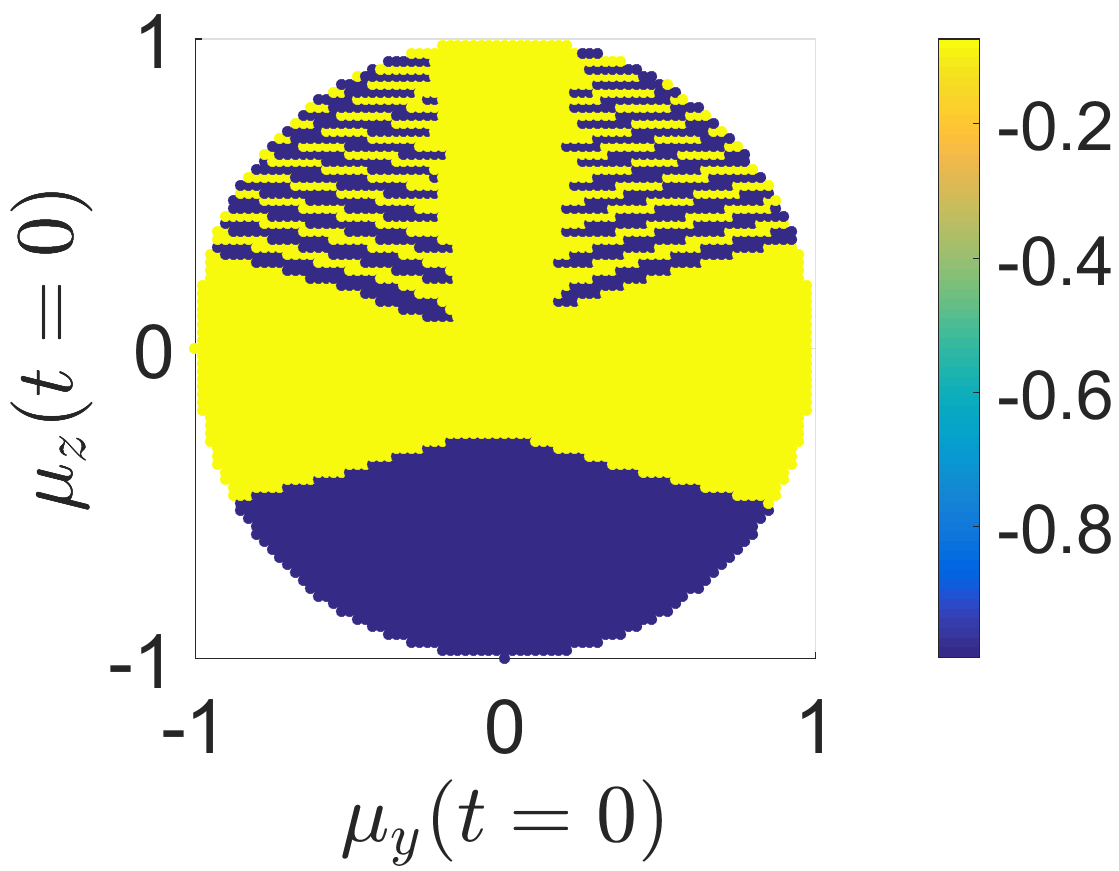}}
\caption{Basins of attraction of the two bistable steady states, depicted by the final $\mu_z^S$ (given by the color code), as a function of the initial condition in a transversal cut through the unit-magnetization sphere, for $\Gamma=1$. The model parameters are increased for each panel, and in panels (a),(b) and (d) the initial condition lies in the plane $\{\mu_y,\mu_z\} $, while in panels (c) the initial condition lies in the plane $\{\mu_x,\mu_z\} $, showing that the choice of the initial plane is not {\it a priori} very restrictive. The basins in the region $\mu_z\ge 0$ are stretched and twisted into each other as the parameters are increased into the weak damping limit $\Omega,\Delta,J\gg\Gamma$. } \label{Fig:Basins}
\end{figure}

\begin{figure}
{\includegraphics[clip,width=0.45\textwidth]{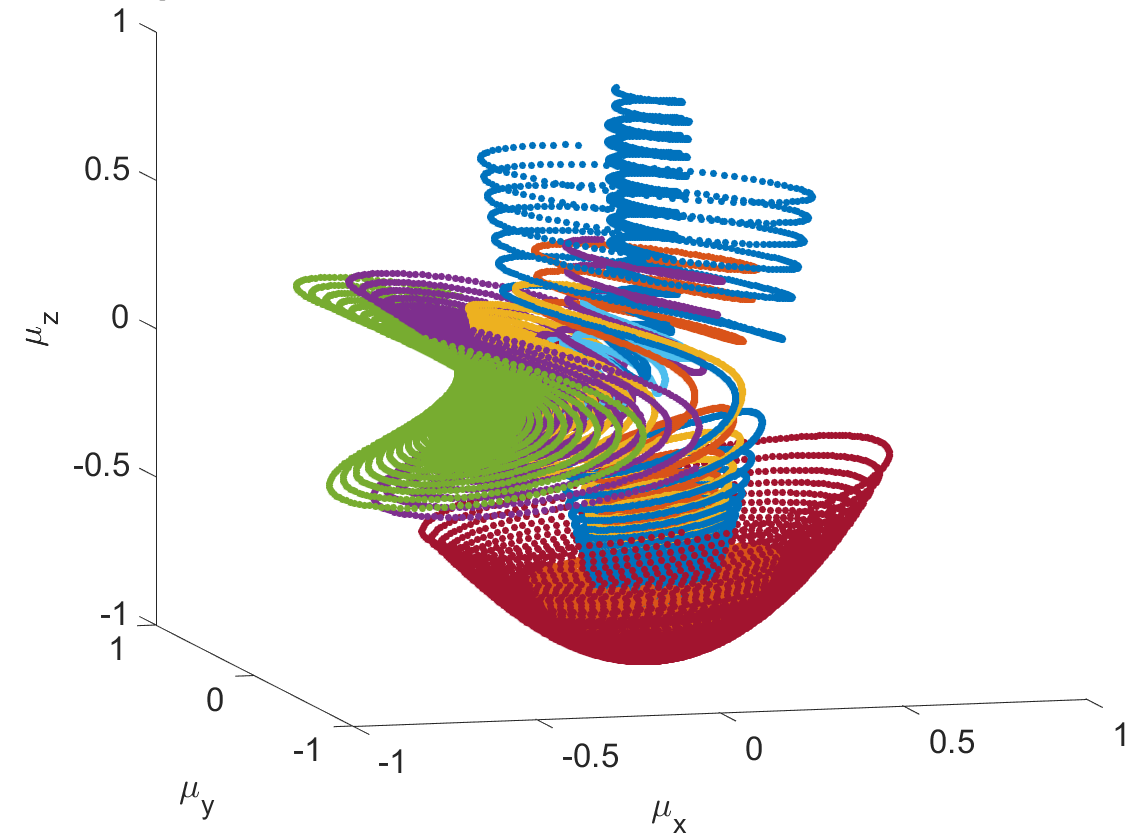}}
\caption{MF trajectories $\vec{\mu}(t)$  for $\Gamma=1$, $\Omega=10$, $J\mathcal{Z}=220$, and $\Delta = 20$, with a few initial conditions. The fast rotations (on the scale of $\Gamma=1$), which are induced by the drive and interactions in the weak damping limit, lead neighboring initial conditions in the upper hemisphere of initial conditions to separate into the two bistable steady states.} \label{Fig:PS}
\end{figure}

In order to visualize the basins of attraction we consider transversal cuts through the state-space, i.e.~by restricting the initial conditions to a plane, e.g.~$\{\mu_x,\mu_z\}$. We find that other planar cuts appear qualitatively similar (as exemplified in \fig{Fig:Basins});
we plot in \fig{Fig:Basins} $\mu_z^S$ (given by the color code), as a function of the initial condition $\vec{\mu}(t=0)$. As can be seen, the basins in the region $\mu_z\ge 0$ are stretched and twisted into each other as the parameters are increased into the weak damping limit (for which $\Omega,\Delta,J\gg\Gamma$).  A quantification of the mixing of the basins of attraction could be done by measuring the length of the boundary curve between the two basins, or perhaps just by counting the number of jumps on the boundary of the circle. A more detailed investigation of the dynamics would be required in order to explain this mechanism. However, \fig{Fig:PS}, showing the dynamics of $\vec\mu$ in the weak damping limit, suggests an initial understanding. It can be seen that many rotations in phase space take place before the solution settles to one of the steady states, with neighbouring initial conditions originating from the upper half of the Bloch sphere separating into the two steady states. Clearly, at 
weak $\Gamma$, the combined effects of fast precession
with the bistability of the final state gives the MF dynamics  some strong sensitivity to the initial condition.

\section{Approaches going beyond mean field}\label{Sec:BeyondMF}
In this section we present two methods allowing to explore the physics of the model beyond the MF approximation.
The first (Sec.~\ref{Sec:MFQF}), MFQF, amounts to dress the MF state at leading order by two-point correlations.
Next, in Sec.~\ref{Sec:MPO}, we describe a numerical method based on MPO which allows for a controlled and accurate approach to the true many-body state in low dimension (1D and thin 2D cylinders). The results obtained by these two complementary techniques will be compared and discussed in Sec.~\ref{Sec:Numerics}.

\subsection{Mean field with Quantum Fluctuations}\label{Sec:MFQF}
Going beyond MF, the next order correction can be included by deriving the e.o.m of $\vartheta_{ab}(R)$ [setting $R'=0$]. 
The approximation we present is based on assuming that $\zeta$, defined in Eq.~\eqref{Eq:tilderhoR0}, and higher order connected correlators, can be neglected in comparison to $\eta$. The e.o.m of $\vartheta_{ab}$ is
\bem  \partial_t\vartheta_{ab}(R)= \sum_d\Pi_{ad}\vartheta_{db}(R) + \sum_d\Pi_{bd}\vartheta_{ad}(R)\\+ f_{ab} (\mu,\vartheta)+ g_{ab} (\mu,\vartheta), \label{Eq:dotetar_ab}\end{multline}
where the local Hamiltonian terms are described using the matrix
\be  \Pi= \left(\begin{array}{ccc}
 0 & -\Delta & 0
  \\
  \Delta & 0 & -2{\Omega} 
  \\
0 & 2{\Omega} &  0
   \end{array}\right),\ee
while $f_{ab}(\mu,\vartheta)$, which contains terms proportional to $J$ and to $J_z$, comes from the kinetic terms, and $g_{ab}(\mu,\vartheta)\propto \Gamma$ comes from the Lindbladian part.  Both are derived in \app{App:eom}. By using \eq{Eq:etadef} we get the e.o.m system for $\eta(R,t)$,
\be \partial_t \eta_{ab}(R,R',t)=\partial_t \vartheta_{ab}(R,R') - \partial_t \left[\mu_a\mu_b\right],\label{Eq:etaeom}\ee
which we solve numerically together with the coupled system for $\vec\mu(t)$ [\eqss{Eq:eomRmag_x}{Eq:eomRmag2}], on lattices of varying sizes, surpassing one hundred thousand sites.

We consider the covariance matrix of the total magnetization,
\bem \left\langle \left[\sum_{R}\left(\sigma_R^a- \mu_a\right)\right] \left[\sum_{R }\left(\sigma_R^b -\mu_b\right)\right]\right\rangle =\\  \left\langle \sum_{R,R'}\left (\sigma_R^a \sigma_{R'}^b-\mu_a \mu_b\right)\right\rangle = \\  N \left(\delta_{a,b} + i\epsilon_{abc} \mu_c-\mu_a \mu_b\right)+N \sum_{R\neq 0}\eta_{ab}(R), \end{multline} whence the imaginary term drops from the symmetrized covariance per spin, which has a finite nontrivial value in the thermodynamic limit $N\to\infty$,
\be \tilde\Sigma_{ab}/N= \left(\delta_{a,b} -\mu_a \mu_b\right)+ \sum_{R\neq 0}\eta_{ab}(R).\label{Eq:Sigma}\ee
The first terms result from the local properties of the spin-one-half system. In the following we will study the total (connected) correlation as a measure of the correlations in a fluctuating domain,
\be \Sigma_{ab}= \sum_{R\neq 0}\eta_{ab}(R).\ee

\subsection{Matrix product operators}\label{Sec:MPO}

We numerically solve the Lindblad equation using an MPO representation of the density matrix of the system~\cite{zwolak_mixed-state_2004,verstraete_matrix_2004,prosen_matrix_2009,benenti_charge_2009,mascarenhas_matrix-product-operator_2015}.
Since the density matrix can be considered as a pure state (i.e. a wave function) in some enlarged
Hilbert space with four states per sites, it can be encoded as an matrix-product state (MPS) in that enlarged space.
In this vectorized representation, the density matrix is often noted $|\rho\rangle\rangle$, as a ``super ket''.
This point of view  allows to implement the Lindblad evolution
in a way that is formally similar to the unitary evolution of a pure state in tDMRG, the Hamiltonian being replaced by the Lindbladian (super)operator.

One qualitative difference with the unitary evolution is of course the fact that the ``norm'' $\langle\langle\rho\rangle\rangle={\rm Tr}\left[\rho^2\right]$ is not conserved during the time evolution.
The latter is simply related to the second R\'enyi entropy of the system, $S_2=-\ln {\rm Tr}\left[\rho^2\right]$.
If we denote by  $|1\rangle\rangle$ the super ket representing the identity density matrix, the scalar product $\langle\langle 1|\rho\rangle\rangle={\rm Tr}\left[\rho\right]=1$ is, however, conserved.
In addition, in presence of dissipation,  a finite-system is expected to have a unique steady-state, independent of the initial conditions (note that this may no longer be true if one first takes the  thermodynamic limit and then the limit of long times~\cite{bimodality}).

As for  MPS-based methods describing pure states, an MPO-based description of a mixed state gets more and more precise as the so-called bond dimension is increased. In 1D, we expect that (in generic situations)
the bond dimension required to achieve a given precision does not grow with the system size (like when encoding a  gapped pure state with MPS). For this reason one can access long times and very accurate results for large 1D systems.

In the same way as MPS methods can be used for 2D lattices, using a snakelike path visiting all sites~\cite{stoudenmire_studying_2012}, one can encode the density matrix of a 2D mixed state using an MPO. One price to pay is the fact that interactions that are local in 2D become long-ranged along the one-dimensional path. For this approach the most natural geometry is that of a cylinder, with open boundary conditions in the $x$ direction, and periodic ones in the $y$ direction. In that case
the  bond dimension required to achieve a given precision grows exponentially with the cylinder diameter $L_y$, contrary to genuinely 2D representations (see for instance \cite{KshetrimayumEtalNatComm2017}). On the other hand, with MPS and MPO one can take advantage of the efficient and the well controlled algorithms that have been developed to evolve and optimize matrix-products objects. As for the $x$ direction, the numerical cost (time and memory) is linear in $L_x$.  The calculations presented here are limited to $L_y=4$, where a bond dimension of the order of a few hundred is enough
to give some good precision.

\begin{figure}
\includegraphics[clip,width=0.48\textwidth]{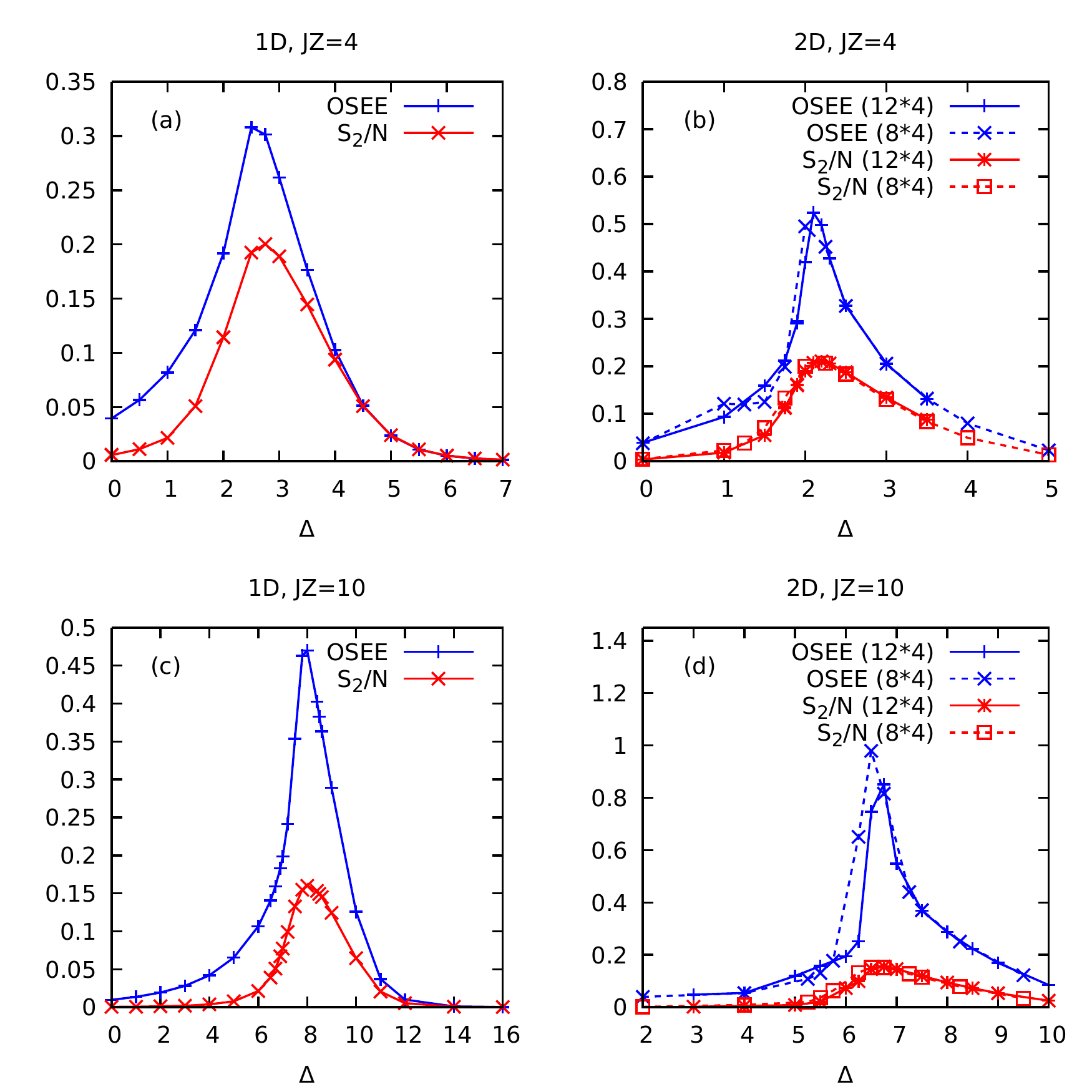}
\caption{Second Rényi entropy per site in the steady state, $S_2/N$ (red), and 
operator-space entanglement entropy (OSEE) associated to the bipartition 
of the system in the center (blue).
These  entropies allow to identify the crossover region where the steady 
state is most
distant from a pure state (large $S_2$), and where it is the most 
correlated (maximum of the OSEE).
The data were obtained by MPO simulations on 1D systems (of size $N=100$ 
or 200) and 2D systems (cylinders with perimeter $L_y=4$ and length 
$L_x=8$ or 12 (full lines or dashed lines)), for $J\mathcal{Z}=4$ and $J\mathcal{Z}=10$.
In 1D the OSEE saturates with the system size, whereas it is expected to 
be proportional to $L_y$ in 2D.
The maximal MPO bond dimension used in the calculations is 200 or 300 in 
1D (depending on the value of $\Delta$) and it is equal to 400 in the 2D 
cylinders.}
\label{Fig:MPOOSEE}
\end{figure}

A quantity of interest is the Von Neumann entanglement entropy associated to the pure state $|\rho\rangle\rangle$. It can be computed for any bi-partition of the system, and is
called the operator space entanglement entropy (OSEE) \cite{prosen_operator_2007}.
For a product state ($\rho=\bigotimes_i \rho_i$), mixed or pure, the OSEE vanishes. For a pure state, the OSEE is twice the usual Von Neumann entropy associated to the same bipartition.
This entropy quantifies the total amount of correlations, classical and/or quantum, between the two subsystems.
It also quantifies how ``demanding'' it is to represent (or approximate) $\rho$ in an MPO form.
\fig{Fig:MPOOSEE} represents the steady state OSEE as a function of $\Delta$, for two values of $J\mathcal{Z}$ at $\Omega=0.5$ and $\Gamma=1$.
The partition considered here corresponds to a left-right cut in the center, with two subsystems of equal sizes.
Both 1D chains and 2D cylinders are considered, and the results allow to identify the interesting range of $\Delta$ where
the steady state is the most correlated, and thus the most distant from a MF product state. In 2D cylinders the OSEE is expected to
be proportional to the perimeter $L_y$, which would be  the analog of the ``area-law'' scaling for the entanglement entropy in pure states. Since the maximal value of the OSEE turns out here to be quite moderate (less than unity), it might be possible to investigate cylinders with a slightly larger perimeter in future studies.

The OSEE is sensitive to all (connected) correlations between the two subsystems, but it does not distinguish between classical and quantum correlations. In the case of pure states the Von Neumann entropy (of a subsystem) is the usual measure for entanglement, and it is specifically sensitive to quantum effects. But the entropy, computed from the reduced density matrix of a subsystem, is generically nonzero in any mixed state, even if the problem is purely classical. In the present model the steady-states are of course not completely classical, but in future studies it would be interesting to quantify the amount of ``quantumness'', that is how far the state is from separable states.

\section{Results beyond mean field}\label{Sec:Numerics}


\subsection{Dimension one}

\begin{figure}
{\includegraphics[clip,width=0.45\textwidth]{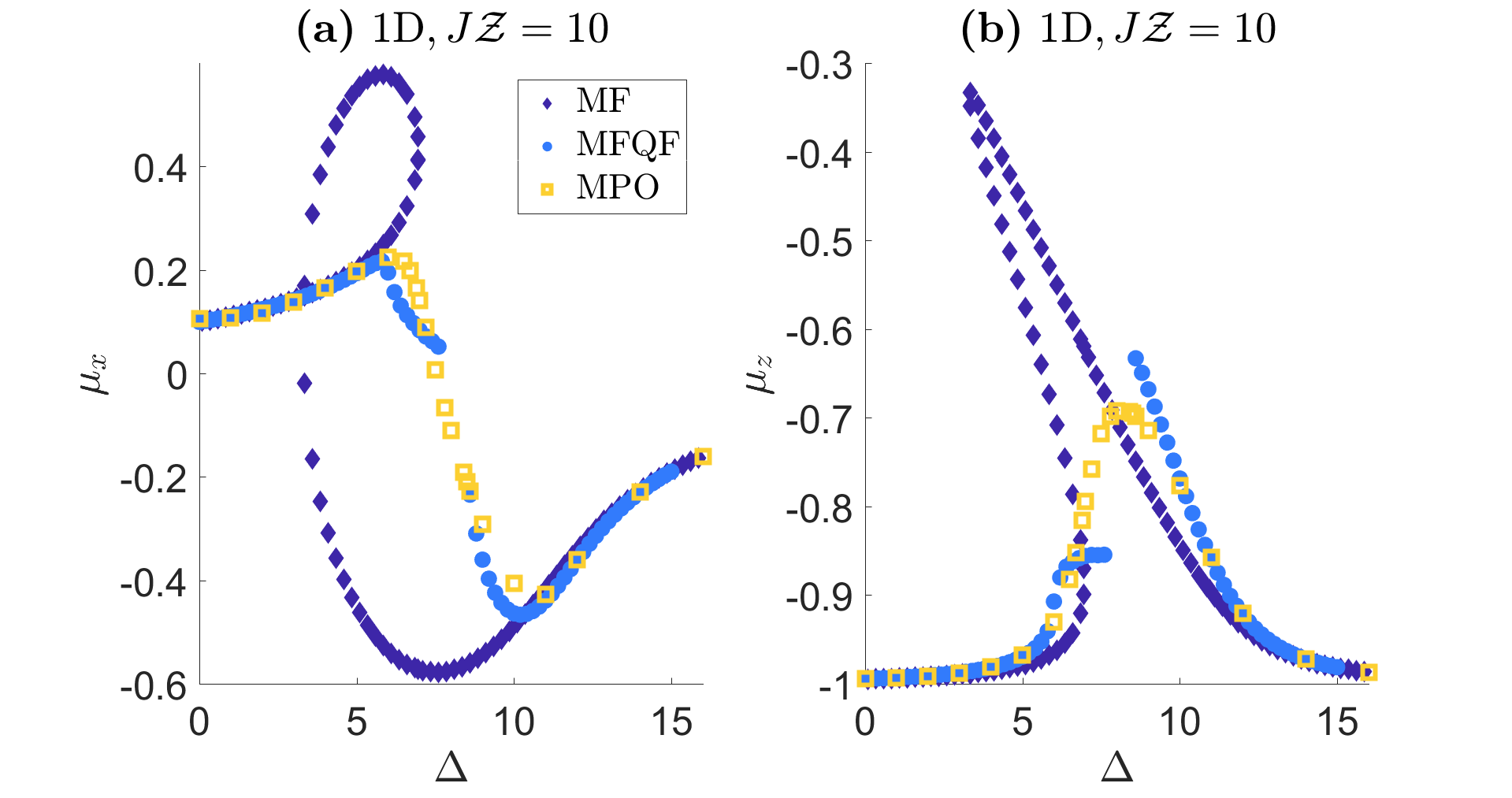}}
\caption{Mean steady state (a) $x$ magnetization $ \mu_x^S$, and (b) $z$ magnetization $ \mu_z^S$, as a function of $\Delta$, for fixed values of the other paramaters, $\Gamma=1$, $\Omega=0.5$, $J=5$, on a 1D chain ($J\mathcal{Z}=10$). The MF limit manifests bistability for $3.4\lesssim\Delta \lesssim 7$. MPO shows that the jump is smoothened to a crossover, within a range of detuning ($6\lesssim\Delta \lesssim 11$) shifted from the MF bistability region. The MFQF approximation results in a unique phase that follows approximately the exact result in some range of $\Delta$. For $7.6\lesssim\Delta \lesssim 8.4$ this approach breaks as the correlations become too large, and no data points are plotted. An analysis of the correlation functions is presented in \fig{Fig:Correlations}.} \label{Fig:Comparison05}
\end{figure}

\begin{figure}
\includegraphics[clip,width=0.47\textwidth]{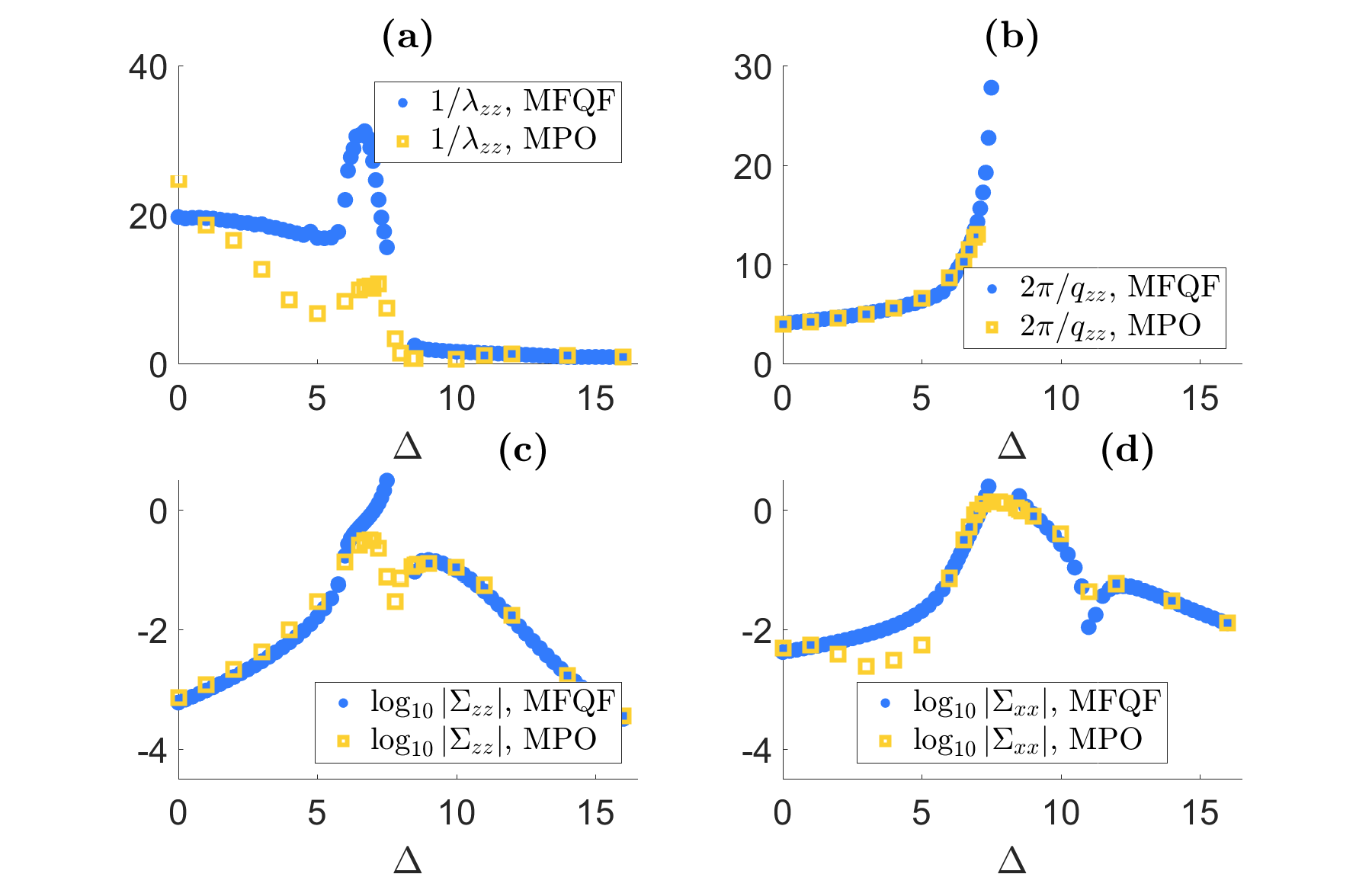}
\caption{Characterization of the two-point correlation functions $\eta_{ab}(R)$ [see \eq{Eq:eta_Rf}] in the steady-state of a 1D system and the comparison of MFQF with MPO, for the same parameters as in \fig{Fig:Comparison05}. (a) The correlation length [$1/\lambda_{ab}$, see \eq{Eq:eta_Rf}] drops sharply in the middle of the crossover region, where also; (b) the spatial period of oscillations [$2\pi/q_{ab}$], shows a sharp increase, beyond which $\eta_{ab}(R)$ are overdamped functions of distance.
(c)-(d) Total correlation $\Sigma_{ab}=\sum_R \eta_{ab}(R)$, showing a increase by up to three orders of magnitude in quantum fluctuations within the crossover region. For low $\Delta$ values, $\eta_{ab}$ are relatively small but spatially extended, while for high $\Delta$ values $\eta_{ab}$ are much larger but short-ranged. The approximate MFQF expansion captures the features of the correlations qualitatively and quantitatively, except at the center of the crossover region, where the correlations are over-estimated by the approximation and in some $\Delta$ range diverge leading to a breakdown of the method.}
\label{Fig:Correlations}
\end{figure}

As discussed in \cite{bimodality} and shown also in \fig{Fig:Comparison}, the difference between the steady state magnetization $\vec\mu^S$ obtained in 1D with MPO
and the MF ellipse increases with the interaction strength $J$. This deviation is induced by the presence of nonzero correlations at distance $=1$ in the lattice, as can be seen from (the exact) \eqss{Eq:eomRmag_x}{Eq:eomRmag2}.
Figure \ref{Fig:Comparison05} compares $\vec\mu^S$ for MF, MFQF and MPO through the crossover region in $\Delta$, for a larger value of $J\mathcal{Z}=10$ in 1D. The corrections to MF
are significant for $6\lesssim\Delta \lesssim 11$. To study the correlation functions we quantify the six independent components of $\eta_{ab}(R)$ by their discrete Fourier transform and the correlation length. \fig{Fig:Correlations} presents two lengthscales -- the correlation length $1/\lambda_{ab}$ and the inverse of the dominant wavevector $q_{ab}$ -- which amounts to a dominant functional dependence on distance in the form \be\eta_{ab}(R) \sim \exp\{-\lambda_{ab}R\}\left[A_{ab}+B_{ab}\cos(q_{ab}R+\phi_{ab})\right],\label{Eq:eta_Rf}\ee
 where $A_{ab}$, $B_{ab}$ and $\phi_{ab}$ are coefficients. In the heart of the crossover region across the MF bistability, the correlations calculated in MFQF or MPO in 1D chains grow by up to a few orders of magnitude, as measured by $\Sigma_{ab}= \sum_R\eta_{ab}(R)$ [see \eq{Eq:Sigma}].
As can be deduced from \fig{Fig:Correlations}, the spatial structure of the two-point correlation functions undergoes a qualitative change within the crossover region.
For low $\Delta$ values the correlations have a relatively small amplitude, but they decay slowly with distance (large correlation length)
and display some incommensurate density-wave character. On the other hand, for high $\Delta$, the correlations are very short-ranged (overdamped in space) and do not exhibit oscillations.  

\begin{figure}
\includegraphics[clip,width=0.37\textwidth]{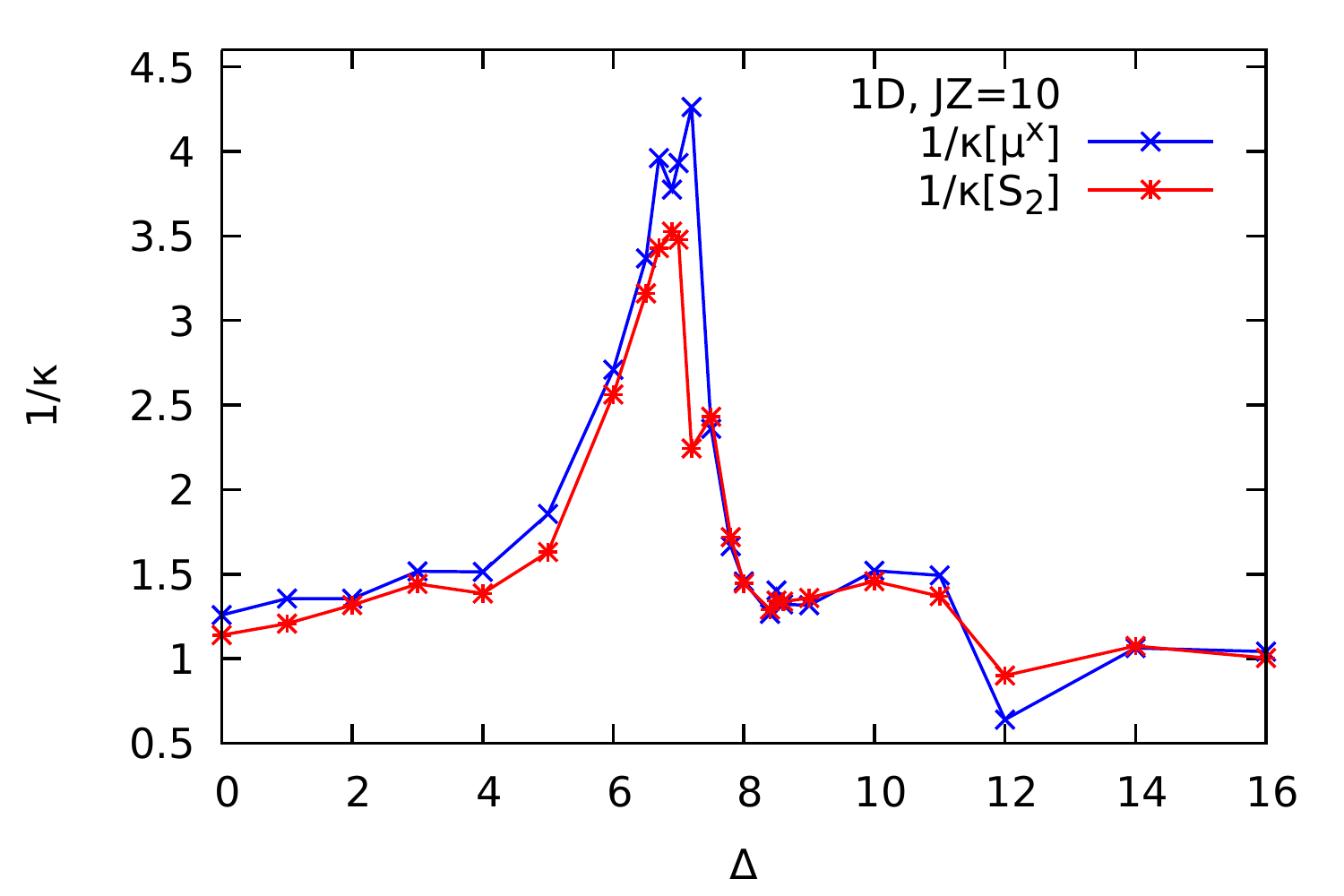}
\caption{Relaxation time $1/\kappa$ extracted from the MPO simulations 1D (same parameters 
as in Fig.~\ref{Fig:Comparison05}).
Two relaxation times are shown: one associated to the relaxation of the 
$x$ component of the magnetization
to its steady value (blue), and the other one associated to the second 
Rényi entropy $S_2$ (red).
Both show a  similar behavior, with a peak (of finite height) in the 
crossover region.}
\label{Fig:MPORelax}
\end{figure}

Although the system does not show any bistability, we observe some enhanced relaxation time in the crossover region, as shown in \fig{Fig:MPORelax}. There, the relaxation to the steady state was fitted to an exponential decay $\sim \exp(-\kappa t)$. The relaxation times $1/\kappa$ associated to the $x$-magnetization as well as that associated to the second Rényi entropy $S_2$ are shown. Although these two quantities are  very different in nature, they give very similar relation times. The rate $\kappa$ extracted from the dynamics of other observables, like $\mu^z$ or the OSEE for instance, also give very similar results. This suggests that we are here probing some intrinsic timescale of the model, proportional the inverse of the Liouvillian gap.


From \fig{Fig:Correlations} it can be seen that the crossover region is well captured by the MFQF approximation.
Although quantitatively overestimated in the crossover region,  the correlation length predicted by MFQF behaves in a way that is qualitatively very similar to that given my MPO calculations [\fig{Fig:Correlations}(a)]. But what is quite remarkable is the agreement observed in \fig{Fig:Correlations}(b) concerning the wave-vector of the modulations of the correlations. It appears that
the MFQF formalism captures almost exactly this incommensurate character of the correlations.
The magnitude of the correlations, probed here via $\Sigma_{zz}$ and $\Sigma_{xx}$ [see Eq.~\eqref{Eq:Sigma}], also behaves in a way that is qualitatively similar to the MPO data. We note, however, that for $7.6\lesssim\Delta \lesssim 8.4$ the MFQF approximation breaks down due to quantum fluctuations becoming too large and the correlators grow to nonphysical ($>1$) values. For this reason, it is in this range that the deviations from the MPO results are the largest.

\begin{figure}
\centerline{\includegraphics[clip,width=0.50\textwidth]{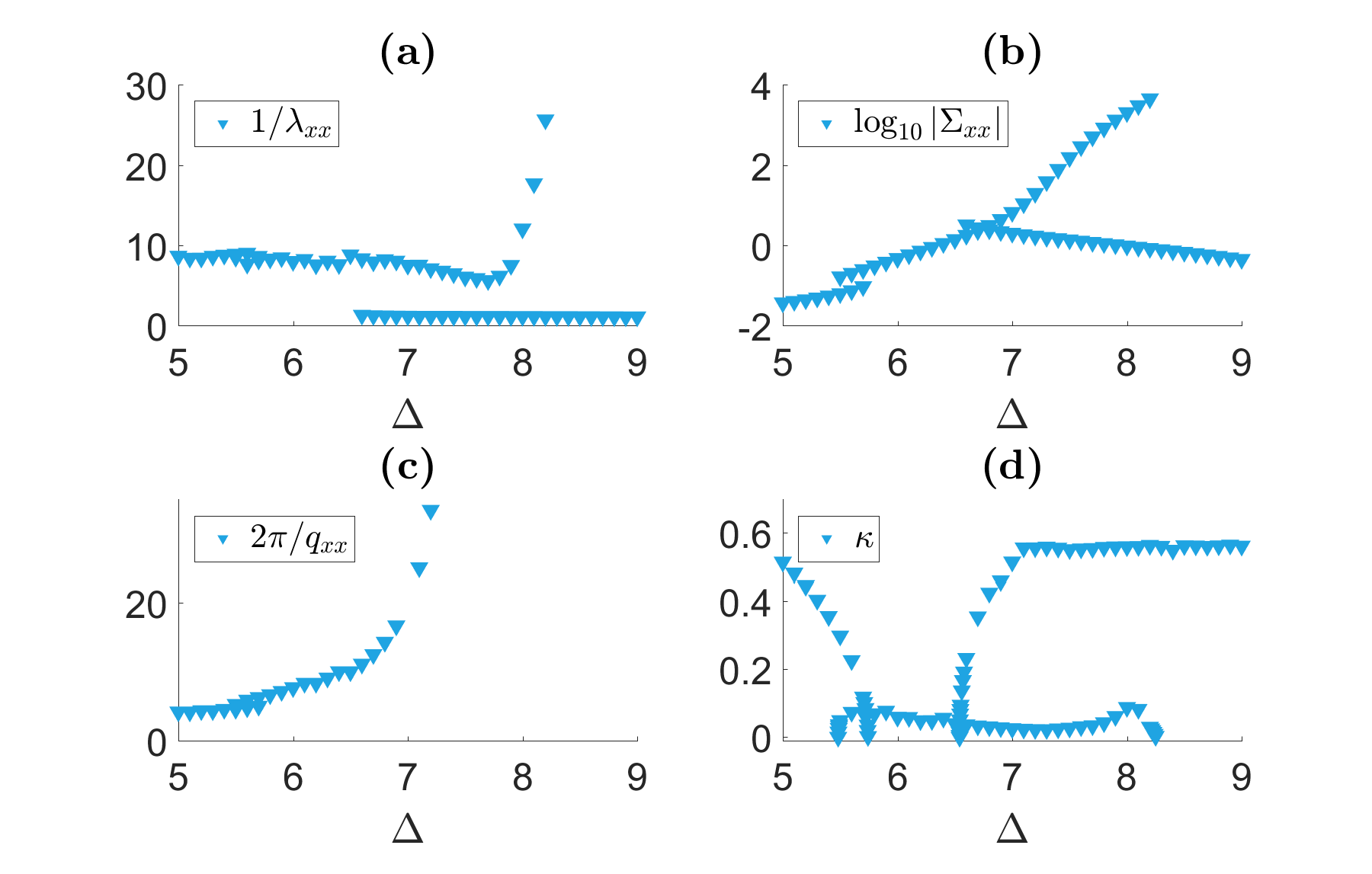}}
\caption{Characterization of the two-point correlation functions $\eta_{ab}(R)$ in 2D, for the same parameters as in \fig{Fig:Comparison052D}, within the MFQF approximation. Three branches of solutions are shown. (a) The correlation length, which diverges at the right edge of the intermediate branch ($\Delta\approx 8.2)$. (b) Total correlation $\Sigma_{ab}=\sum_R \eta_{ab}(R)$, showing a large increase in quantum fluctuations within the intermediate branch. (c) The spatial period of oscillations shows a sharp increase in the intermediate branch, beyond which $\eta_{ab}(R)$ are overdamped functions of distance.
(d) The rate of convergence to the steady state, taking the exponential form $\sim\exp\{-\kappa t\}$, showing a critical slowing down of the decay dynamics at the edges of the phase branches. See text for details.}
\label{Fig:Correlations2D}
\end{figure}

	\subsection{Higher dimensions}

\begin{figure}
\includegraphics[clip,width=0.48\textwidth]{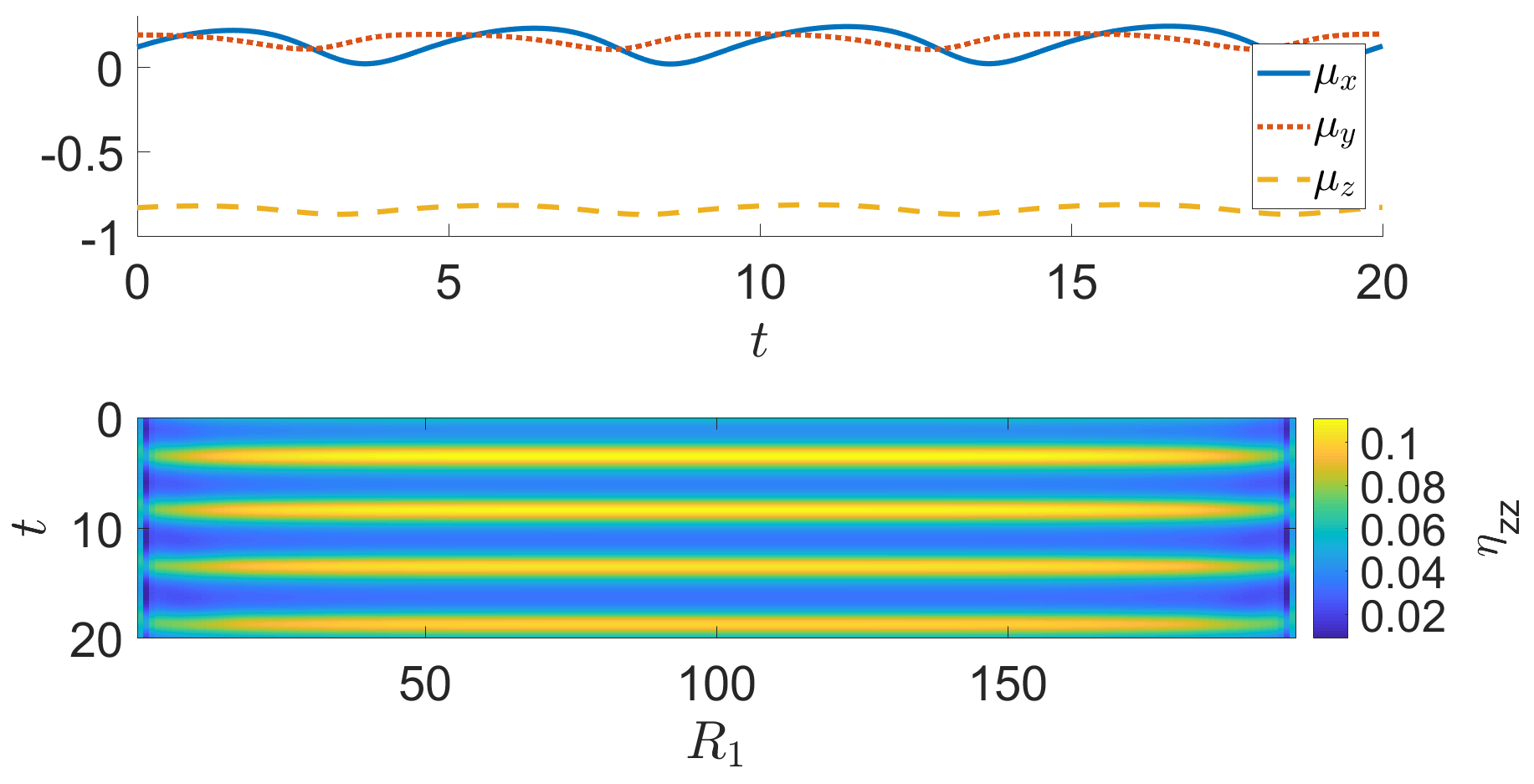}
\caption{{\it Upper panel --} The oscillations of the MFQF magnetization in the limit cycle state in 2D for $\Delta=8.3$ with the other parameters as in \fig{Fig:Comparison052D} [$\Gamma=1$, $\Omega=0.5$, $J\mathcal{Z}=10$]. {\it Lower panel --} A space-time diagram of the 2D correlation function $\eta_{zz}(R,t)$, with $R=\{R_1,R_2=0\}$ taken along a 1D cut through the lattice. Nondecaying correlations extending over the simulated system size (with $200^2$ sites), manifest oscillations (notable at short-range distance) that are coupled to the magnetization oscillations.}
\label{Fig:LC}
\end{figure}

As discussed in the introduction and in \cite{bimodality}, in dimension two and higher, MFQF predicts bistability.
While in 2D our MPO simulations could be used to benchmark the MFQF results in regions where the correlation length is not too large, the accessible system size are relatively small, resulting necessarily in a unique steady state.

We therefore focus here on the MFQF results on 2D lattices, with the same parameters as the 1D example discussed above, with $J\mathcal{Z}=10$ fixed (hence $J=2.5$). The MFQF approximation converges throughout the $\Delta$ range presented in \fig{Fig:Comparison052D}, and the maximal values taken by the correlation functions are smaller than in 1D. Figure \ref{Fig:Correlations2D} shows the characteristics of the correlation functions for the three steady-state branches.
We find that the new emerging branch at intermediate magnetization values shares some properties
with the MF-like branch that is stable at low $\Delta$ values: a large correlation length and  spatial oscillations characterized by $q_{ab}\neq 0$. 
At the same time, the new branch has large 
absolute values of the correlation functions, as is the case with the MF-like branch that is stable at $\Delta\gg 1$.

The inverse of the relaxation time to the steady state, $\kappa$,  is plotted in \fig{Fig:Correlations2D}(d).
It is obtained by fitting $\partial_t \vec\mu^2$, the time derivative  of the square of the magnetization, to $\sim\exp\{-\kappa t\}$ at large times (of order $t\sim 100$). $\kappa$ thus measures
decay rate of small perturbations about the steady state.
It can be seen that
$\kappa$ vanishes at the end points of each branch. In 2D we thus observe some critical slowing down at the edge of each branch (this holds in the MF approximation as well). Some important slowing down is also observed in the MPO calculations performed on cylinders, although the relaxation time cannot diverge on the small systems we considered.

As discussed briefly in \seq{Sec:MainResults}, we find that the new steady state and the limit cycle (LC) are stable in smaller $\Delta$ regions as the dimension is increased. As with the LC, an exact characterization of this dependence is beyond the scope of the current work (and it becomes increasingly demanding to study the dependence of the results on $N$ as the dimension is increased). 

For $\Delta\approx 8.24$ (see \fig{Fig:Correlations2D}), the correlation length in the new branch diverges, and beyond this point we find a stable LC, coexisting together with the MF-like branch in a small $\Delta$ range (up to $\Delta\approx 8.32$). The amplitude of the magnetization oscillations varies with $\Delta$ within the range of stability of the LC, and so does its frequency and other characteristics. In \fig{Fig:LC} we illustrate the LC phenomenon,
showing oscillations of the mean magnetization as a function of time, and the space-time pattern in the correlation functions (along one spatial coordinate in the square lattice). In particular, the connected two-point function $\eta_{zz}$ becomes spatially long-ranged at periodic intervals in time [yellow lines in   \fig{Fig:LC}].

\section{Summary and Outlook}\label{Sec:Outlook}

We have presented a detailed characterization of the MF limit of the driven-dissipative XY spin model, and studied some aspects of the associated dynamics such as the basins of attraction of different steady states. Going beyond MF, we have employed some accurate MPO-based method and the approximate MFQF approach. We have addressed the existence of multistability of the steady state predicted in MF, in addition to the possibility of states not captured by the MF limit, together with the spontaneous emergence of long-range spatial and temporal order. 

As we have seen, in 1D the MFQF approach is capable of converging to a unique steady state, giving a picture which is very different from MF. As a comparison to MPO simulations shows, it is also capable to qualitatively and quantitatively predict some characteristics of the correlation functions, except in parameter regimes where the correlations become so large such that the neglect of three-point correlations (and higher) renders the approximation nonconverging. As the dimension is increased the correlations decrease in absolute magnitude. There, the MFQF is expected to become more accurate and this has been confirmed by comparison to MPO simulations of small 2D systems, for parameters that allow the comparison. It is straightforward to include in the MFQF formalism some more general local Lindblad terms acting on the spins, like  the effect of dephasing. This could be a first immediate extension of the current work. Also, the investigation of other lattice geometries and/or longer-ranged spin 
interactions (as in \cite{PhysRevA.85.043620,PhysRevA.89.023616,parmee_steady_2019}), all of which are relevant for 2D systems of trapped ions, constitute interesting future directions.

The treatment of correlations in MFQF can account for  two-point correlations with a large correlation length. This has allowed us to find a new steady state branch in 2D and higher, which is lacking in the MF limit.
Clearly, the regimes with a large correlation length would be difficult to capture using methods based on small clusters. This intermediate branch shares some of its characteristics with the two MF-like steady states, making it suggestive to speculate that it may consist pictorially of coexisting domains. The correlation length in this branch diverges at some point, which may be an indication that the true quantum solution tends to develop long-range order in the form of a superposition. Beyond this critical point, MFQF predicts a correlation-induced limit cycle state, with correlation functions again extending over the simulated  system size, and hence compatible with long-range spatio-temporal order. It should be noted that the MFQF approximation allows simulating very large lattices, but of course, it remains an open question how the inclusion of higher-order correlations would affect this 
behaviour. 

The application of MPO simulations to 2D lattices, demonstrated here for the first time for a system with Lindblad dynamics, has proven successful. The numerical cost of such simulations is exponential in the perimeter ($L_y$) of the cylinder, but only linear in its length. This allows one to compute accurately in a controlled way not only the steady state properties, but also the transient dynamics of the system, for systems that are significantly larger than what is doable with an exact brute-force approach, while still keeping all the many-body correlations. In the present study we have been deliberately conservative, in the sense that we kept the systems sufficiently small so that the MPO calculations were essentially exact. These MPO calculations on small cylinders could then be used to benchmark the MFQF results in situations where the correlation length was small enough. It would of course be very interesting to push the 2D MPO simulations further in order to see if one can reach big enough systems and confirm the new phenomena predicted to occur in 2D by the MFQF approach. 

Finally, we have shown that the basins of attraction of the new steady state progressively decrease with the dimension. We have also found that the parameter range of stability of the limit cycle again decreases with the dimension, possibly disappearing above 4D. Although it is hard to verify that these observations are independent of the simulated lattice sizes (due to the lateral lattice dimension accessible in our simulations necessarily decreasing with $D$, while the correlation length much less, or not at all for some parameters), it provides a plausible explanation to the fact that these new phases do not survive in the MF limit.
The possibility of simulating these spin models with controllable parameters using systems of trapped ions in 1D \cite{brydges2019probing,smith2016many, ramos2019feasibility} and 2D \cite{bohnet2016quantum}, and arrays of superconducting qubits \cite{AndrewNatPhys,SchmidtKochAnnPhy13, LeHurReview16,HartmannJOpt2016,mckay2018qiskit, alexander2020qiskit}, holds a promise for exploring the emergence of phases with long spatial and temporal order out of the competition of coherent driving, dissipation, and strong interactions.

\begin{acknowledgments}
We acknowledge the DRF of CEA for providing us with CPU time on the supercomputer COBALT of the CCRT. H.L. thanks Roni Geffen for fruitful discussions, and acknowledges support by IRS-IQUPS of Universit\'e Paris-Saclay and by LabEx PALM under grant number ANR-10-LABX-0039-PALM.
\end{acknowledgments}

\appendix

\begin{widetext}

\section{Further properties of the MF steady state}\label{App:MFApp}

To understand the nature of the solutions as a function of all the parameters, we start by considering some simple limits. 
Using \eq{Eq:mu_z_linear} we find that for $\Omega>0$,
\be    \mu_y^S < \frac{1}{2\Omega /\Gamma} \label{Eq:mu_y_ineq}.\ee
In the limit $\Omega \to 0$, the plane of \eq{Eq:mu_z_linear} is nearly horizontal in the $\vec\mu$-space and $\mu_z^S\to -1$, while for $|\Omega|\to\infty$ it is nearly vertical (see \fig{Fig:Traj}). In the latter case, combining \eq{Eq:mu_y_ineq} and the steady-state relation for $\mu_x^S$,
\be \mu_x^S=-2\mu_y^S\left(\Gamma \Delta-\Gamma J\mathcal{Z} + 2J\mathcal{Z} \Omega \mu_y^S\right)/\Gamma^2,\label{Eq:mu_x_quad}\ee
implies that for  $|\Omega|\to\infty$ (with the other parameters fixed), $\mu_x^S,\mu_y^S\to 0 $ and $\mu_z^S\to 0$ or  $\mu_z^S\to -1$ (we have found that in all cases that we study, for $|\Omega|\to\infty$ the steady state is the infinite temperature state, $\vec\mu^S\to \vec 0$).

Fixing the value of the magnetization (within the constraints above), we can solve the steady state equations to
obtain the model parameters $\Omega/\Gamma$ and $\Delta$. If $\mu_y^S=0$, we get $\mu_x^S=0$ and $\mu_z^S=-1$, and this requires $\Omega=0$. For $\mu_y^S\neq 0$ we obtain 
\be \Omega/\Gamma=(1+\mu_z^S)/2\mu_y^S, \quad \Delta=-\Gamma\mu_x^S/2\mu_y^S- J\mathcal{Z} \mu_z^S.\label{Eq:parametersvsMag}\ee
Hence,  at fixed $\Gamma=1$, $\Omega$ is uniquely determined and we get a straight line in $(J,\Delta)$ plane.
Using \eq{Eq:parametersvsMag} we find that the magnetization 
\be \vec{\mu}_u\equiv(\mu_x=1/2,\mu_y=1/2,\mu_z=-1/2),\ee occurs for $\Omega=1/2$ along the line \be \Delta_u=-\Gamma/2+J_u\mathcal{Z}/2.\ee
 The point $c_u\equiv (J\mathcal{Z}=2,\Delta/\Gamma=1/2)$ is a critical point of the bistability region for $|\Omega|/\Gamma=1/2$. This is the only critical point at $J\mathcal{Z}=2$, as exemplified in \fig{Fig:Bistability}. The magnetization $\vec{\mu}_u$ is the unstable steady state solution along the line $\Delta_u(J_u)$ that crosses the bistable region.

\section{Equations of Motion}\label{App:eom}

For spin one-half operators (Pauli matrices), we have the commutation relations, with $a,b,c=\{x,y,z\}$, 
\be
\left[\sigma^{a}_R,\sigma^{b}_{R'}\right]=2i\eps_{abc}\sigma^{c}_R\,\delta_{R,R'},\qquad (\sigma_R^a)^2=1,\label{Eq:sigmaRRcommutations} 
\ee
and the algebra of the ladder operators reads
\bea
 \left[\sigma^{+}_R,\sigma^{-}_{R'}\right]= \delta_{R,R'}\sigma^z_{R},\qquad
 \left[\sigma^{\pm}_R,\sigma^{z}_{R'}\right]=\mp 2\delta_{R,R'}\sigma^{\pm}_R.
\eea
For the local Hamiltonian terms we get
\be \left[h,\sigma_R^+\right]=-\Omega\sigma_R^z+{\Delta}\sigma_R^+, 
\qquad 
\left[h,\sigma_R^-\right] = \Omega\sigma_R^z-{\Delta}\sigma_R^-,\qquad
\left[h,\sigma_R^z\right] = -2\Omega (\sigma_R^+ -\sigma_R^-).\ee
With the anti-commutation relations on the same site, $\left\{\sigma_R^a,\sigma_R^b\right\}=2\delta_{a,b}$, we get the known relation
\be \sigma_R^a \sigma_{R}^b=\delta_{a,b} + i\epsilon_{abc} \sigma_R^c \label{Eq:AnticommutationR},\ee
that allows to simplify the dissipator terms when deriving the e.o.m.

The Hamiltonian part of the e.o.m for $\vartheta$ [\eq{Eq:dotetar_ab}], can be derived most simply from the relation 
\be \left.\partial_t\vartheta_{ab}(R)\right|_{\Gamma=0}=\left\langle\left( \partial_t\sigma_{R}^a\right) \sigma_{0}^b\right\rangle +\left\langle \sigma_{R}^a \left(\partial_t\sigma_{0}^b\right) \right\rangle  ,\ee
and is obtained by multiplying \eqss{Eq:eomsigmaRx}{Eq:eomsigmaRz} by the required operator, and taking the expectation value.
To derive the components of $f (\mu,\vartheta)$ in \eq{Eq:dotetar_ab} from \eqss{Eq:eomsigmaRx}{Eq:eomsigmaRz}, we multiply the e.o.m of $\sigma_R^a$ (with $R\neq 0$) on the right by $\sigma_{0}^b$ and expand the following series
\be \sum_{\substack{
R'\\ \|R'- R\|=1}} \sigma_{R'}^c \sigma_R^d\sigma_{0}^b= \sigma_{0}^c\sigma_{0} ^b \sigma_R^d \delta_{\|R\|,1} + \sum_{\substack{
R'\\\| R'- R\|=1, R'\neq 0}} \sigma_{R'}^c \sigma_R^d\sigma_{0}^b.\label{Eq:series3}\ee
By assuming $\zeta\approx 0$ for the three-point connected correlator of \eq{Eq:tilderhoR0} [with $R\neq R'\neq R''$],
\bem \zeta_{abc}(R,R',R'') = \\\qquad \left\langle \left(\sigma_{R}^a -\mu_a\right)\left(\sigma_{R'}^b-\mu_b\right)\left(\sigma_{R''}^c-\mu_c\right)\right\rangle= \left\langle \sigma_{R}^a \sigma_{R'}^b\sigma_{R''}^c\right\rangle +2 \mu_a\mu_b\mu_c -\mu_a \vartheta_{bc}(R'-R'') -\mu_b \vartheta_{ac}(R-R'') -\mu_c \vartheta_{ab}(R-R'),\label{Eq:tilderhoR}
\end{multline}
and using \eq{Eq:AnticommutationR} we can simplify \eq{Eq:series3} to get
\bem \sum_{\substack{
R'\\ \|R'- R\|=1}} \langle\sigma_{R'}^c \sigma_R^d \sigma_{0}^b\rangle\approx \left[\delta_{c,b}\mu_d + i\epsilon_{cbe} \vartheta_{ed}(R) \right]\delta_{\|R\|,1} -\sum_{\substack{
R'\neq 0\\\| R'- R\|=1}}\left[2 \mu_b\mu_c\mu_d -\mu_b \vartheta_{cd}(R'-R) -\mu_c \vartheta_{bd}(R) -\mu_d \vartheta_{bc}(R')\right]\\
=\left[\delta_{c,b}\mu_d + i\epsilon_{cbe} \vartheta_{ed}(R) \right]\delta_{\|R\|,1} -\left[2 \mu_b\mu_c\mu_d -\mu_b \vartheta_{cd}(1) -\mu_c \vartheta_{bd}(R) \right]\left[\mathcal{Z}-\delta_{\|R\|,1}\right] + \sum_{\substack{
R'\neq 0\\\| R'- R\|=1}} \mu_d \vartheta_{bc}(R').\label{Eq:sigma3series1}\end{multline}
Multiplying on the left the e.o.m of $\sigma_0^b$ by $\sigma_{R}^a$ with $R\neq 0$,
\be \sum_{\| R'\|=1} \sigma_{R}^a\sigma_{R'}^c \sigma_0^d= \sigma_{R}^a\sigma_{R}^c \sigma_0^d\delta_{\|R\|,1} + \sum_{\substack{\| R'\|=1\\  R'\neq R}} \sigma_{R}^a\sigma_{R'}^c \sigma_0^d,\ee
expanding we get 
\bem \sum_{\| R'\|=1}\langle \sigma_{R}^a\sigma_{R'}^c \sigma_0^d\rangle\approx \left[ \delta_{a,c}\mu_d + i\epsilon_{ace} \vartheta_{ed}(R)\right] \delta_{\|R\|,1} -\sum_{\substack{\| R'\|=1\\ R'\neq R}}\left[2 \mu_a\mu_c\mu_d -\mu_a \vartheta_{cd}(R') -\mu_c \vartheta_{ad}(R) -\mu_d \vartheta_{ac}(R-R')\right] \\
= \left[ \delta_{a,c}\mu_d + i\epsilon_{ace} \vartheta_{ed}(R)\right] \delta_{\|R\|,1} - \left[2 \mu_a\mu_c\mu_d -\mu_a \vartheta_{cd}(1) -\mu_c \vartheta_{ad}(R) \right]\left[\mathcal{Z}-\delta_{\|R\|,1}\right] +  \sum_{\substack{
R'\neq 0\\\| R'- R\|=1}} \mu_d \vartheta_{ac}(R').\label{Eq:sigma3series2}\end{multline}

Let us derive $f_{ab} (\mu,\vartheta)$ of \eq{Eq:dotetar_ab} separately for the two cases $J_z=0$ and $J=0$ (of course by linearity, they can be added). For $J_z=0$, multiplying \eqss{Eq:sigma3series1}{Eq:sigma3series2} by $J$ with the correct sign and summing we get, 
\be
f_{xx}(R)= 2J \left[2 \mu_x\mu_y\mu_z -  \mu_x \vartheta_{yz}(1) -\mu_y \vartheta_{xz}(R)\right]\left[\mathcal{Z}-\delta_{\|R\|,1}\right] - 2J\sum_{\substack{
R'\neq 0 \\ \| R'- R\|=1}}\mu_z \vartheta_{xy}(R'),
\ee
\be
f_{yy}(R)= -2J \left[2 \mu_x\mu_y\mu_z -  \mu_y \vartheta_{xz}(1) -\mu_x \vartheta_{yz}(R)\right]\left[\mathcal{Z}-\delta_{\|R\|,1}\right] + 2J\sum_{\substack{
R'\neq 0 \\ \| R'- R\|=1}}\mu_z \vartheta_{xy}(R') ,
\ee
\be
f_{zz}(R)=  -2J \left[ \mu_x \vartheta_{yz}(R) -\mu_y\vartheta_{xz}(R) \right]\left[\mathcal{Z}-\delta_{\|R\|,1}\right] - 2J\sum_{\substack{
R'\neq 0\\\| R'- R\|=1}}\left[ \mu_y \vartheta_{xz}(R')- \mu_x \vartheta_{yz}(R')\right] .
\ee
\bem
f_{xy}(R)= J  \left[ 2\mu_y^2 \mu_z-2 \mu_x^2\mu_z -\mu_y \vartheta_{yz}(1) -\mu_y \vartheta_{yz}(R) + \mu_x \vartheta_{xz}(1) +\mu_x \vartheta_{xz}(R)\right]\left[\mathcal{Z}-\delta_{\|R\|,1}\right] \\ - J\sum_{\substack{
R'\neq 0\\\| R'- R\|=1}}\left[ \mu_z \vartheta_{yy}(R')  -\mu_z \vartheta_{xx}(R')\right],
\end{multline}
\bem
f_{xz}(R)= -J\mu_y\delta_{\|R\|,1} + J \left[ 2 \mu_z^2 \mu_y -\mu_z \vartheta_{yz}(1) -\mu_y \vartheta_{zz}(R) -\mu_x \vartheta_{xy}(R) +\mu_y \vartheta_{xx}(R) \right]\left[\mathcal{Z}-\delta_{\|R\|,1}\right] \\ - J\sum_{\substack{
R'\neq 0\\\| R'- R\|=1}}\left[ \mu_z \vartheta_{yz}(R') + \mu_y \vartheta_{xx}(R') - \mu_x\vartheta_{xy}(R') \right],
\end{multline}
\bem
f_{yz}(R)= J\mu_x \delta_{\|R\|,1} + J \left[ -2 \mu_z^2 \mu_x +\mu_z \vartheta_{xz}(1) +\mu_x \vartheta_{zz}(R) -\mu_x \vartheta_{yy}(R)+\mu_y \vartheta_{xy}(R)\right] \left[\mathcal{Z}-\delta_{\|R\|,1}\right] \\ + J\sum_{\substack{
R'\neq 0\\\| R'- R\|=1}}\left[ \mu_z \vartheta_{xz}(R') -\mu_y \vartheta_{xy}(R') + \mu_x\vartheta_{yy}(R') \right].
\end{multline}

For $J=0$, multiplying \eqss{Eq:sigma3series1}{Eq:sigma3series2} by $J_z$ with the correct sign and summing we get, 
\be
f_{xx}(R)= -2J_z \left[2 \mu_x\mu_y\mu_z -  \mu_x \vartheta_{yz}(1) -\mu_z \vartheta_{xy}(R)\right]\left[\mathcal{Z}-\delta_{\|R\|,1}\right] + 2J_z \sum_{\substack{
R'\neq 0 \\ \| R'- R\|=1}}\mu_y \vartheta_{xz}(R'),
\ee
\be
f_{yy}(R)= 2J_z \left[2 \mu_x\mu_y\mu_z -  \mu_y \vartheta_{xz}(1) -\mu_z \vartheta_{xy}(R)\right]\left[\mathcal{Z}-\delta_{\|R\|,1}\right] - 2J_z \sum_{\substack{
R'\neq 0 \\ \| R'- R\|=1}}\mu_x \vartheta_{yz}(R') ,
\ee
\be
f_{zz}(R)=  0 .
\ee
\bem
f_{xy}(R)= J_z  \left[ 2\mu_x^2 \mu_z-2 \mu_y^2\mu_z -\mu_x \vartheta_{xz}(1) -\mu_z \vartheta_{xx}(R) + \mu_y \vartheta_{yz}(1) +\mu_z \vartheta_{yy}(R)\right]\left[\mathcal{Z}-\delta_{\|R\|,1}\right] \\ + J_z \sum_{\substack{
R'\neq 0\\\| R'- R\|=1}}\left[ \mu_y \vartheta_{yz}(R')  -\mu_x \vartheta_{xz}(R')\right],
\end{multline}
\be
f_{xz}(R)= J_z\mu_y\delta_{\|R\|,1} - J_z \left[ 2 \mu_z^2 \mu_y -\mu_z \vartheta_{yz}(1) -\mu_z \vartheta_{yz}(R) \right]\left[\mathcal{Z}-\delta_{\|R\|,1}\right] \\ + J_z\sum_{\substack{
R'\neq 0\\\| R'- R\|=1}}\mu_y \vartheta_{zz}(R'),
\ee
\be
f_{yz}(R)= -J_z\mu_x \delta_{\|R\|,1} + J_z \left[ 2 \mu_z^2 \mu_x -\mu_z \vartheta_{xz}(1) -\mu_z \vartheta_{xz}(R)\right] \left[\mathcal{Z}-\delta_{\|R\|,1}\right] \\ - J_z\sum_{\substack{
R'\neq 0\\\| R'- R\|=1}}\mu_x \vartheta_{zz}(R') .
\ee

The components of $g (\mu,\vartheta)$ in \eq{Eq:dotetar_ab} are given by
\be g_{aa} = -\Gamma\vartheta_{aa},\quad g_{xy}=-\Gamma\vartheta_{xy},\quad g_{xz}=-\Gamma\left[2\vartheta_{xz}+\frac{3}{2}\mu_x \right],\quad g_{yz}=-\Gamma\left[ 2\vartheta_{yz}+\frac{3}{2}\mu_y \right].\ee
The spin length evolves according to
\be \partial_t\vec{\mu}^2\equiv \partial_t\left(\mu_x^2+\mu_y^2+\mu_z^2\right)=2J\mathcal{Z}\left[\mu_y\eta_{xz}(1)-\mu_x \eta_{yz}(1)\right]-\Gamma\left( \vec{\mu}^2+ \mu_z^2 + 2\mu_z\right).\label{Eq:dmudteta}\ee

\end{widetext}

\section{Convergence of the MFQF method}\label{App:ConvMFQF}

\begin{figure}
\includegraphics[width=\linewidth]{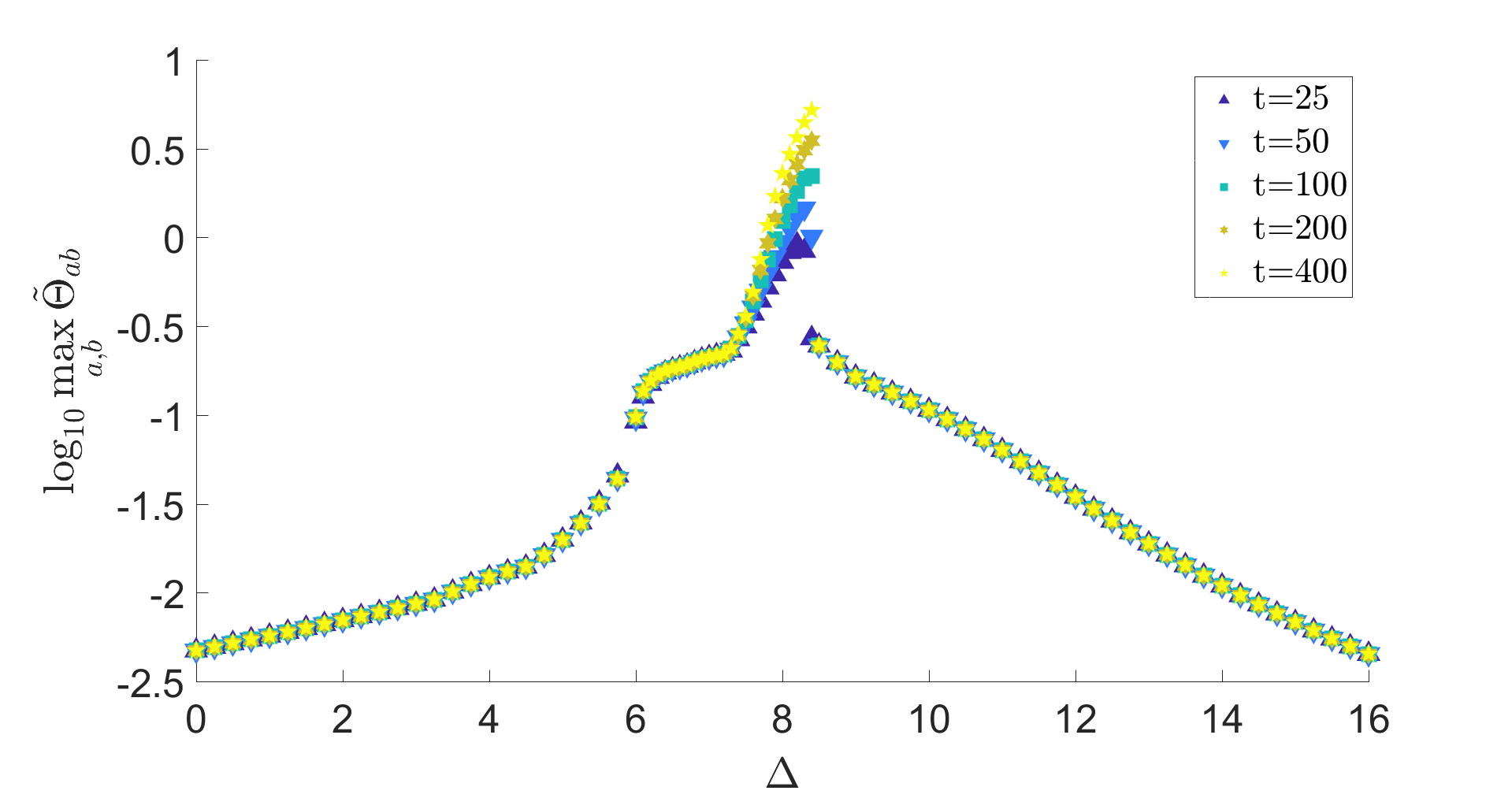}
\caption{The measure of convergence of the correlation functions used for MFQF [\eq{Eq:tildelambda}], for $\Gamma=1$, $\Omega=0.5$, $J\mathcal{Z}=10$ on a 1D chain with $N=1000$ sites.}
\label{Fig:tildelabmda1}
\end{figure}

\begin{figure}
\includegraphics[width=\linewidth]{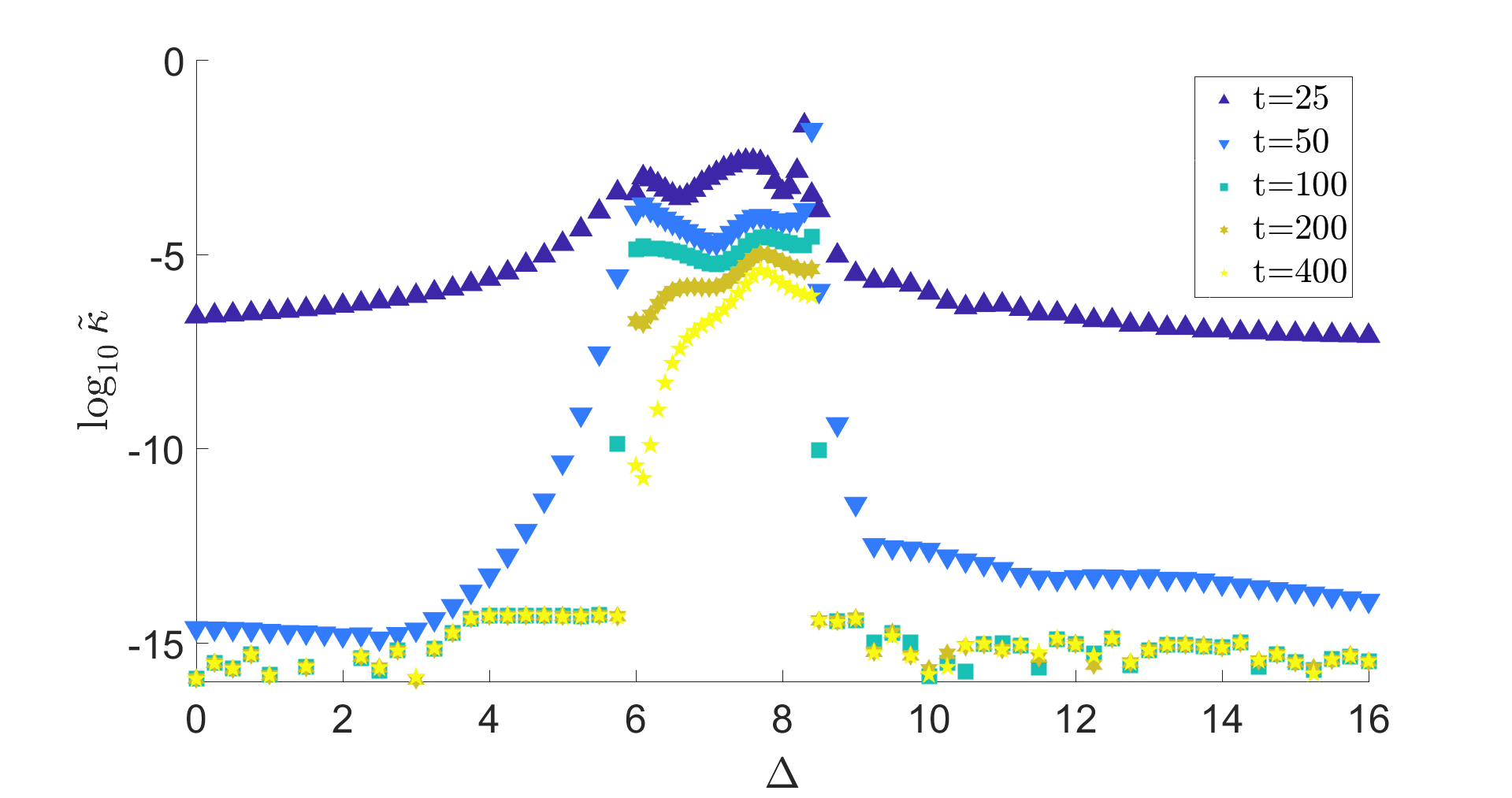}
\caption{The measure of convergence of the time dynamics used for MFQF [\eq{Eq:tildekappa}], for $\Gamma=1$, $\Omega=0.5$, $J\mathcal{Z}=10$ on a 1D chain with $N=1000$ sites.}
\label{Fig:tildekappa1}
\end{figure}

As a measure of the maximal correlations at a given time $t_0$ we take
\be \tilde\Theta_{ab}(t_0)=\max_{t\in[t_0-T,t_0]}\max_R|\eta_{ab}(R)|.\label{Eq:tildelambda}\ee
where $T$ is a small averaging window.
As a measure of the convergence of the dynamics we define using \eq{Eq:dmudteta} in a similar interval,
\be \tilde\kappa(t_0)=\frac{1}{T}\int_{t_0-T}^{t_0}|{\partial_t\vec\mu^2}|dt,\label{Eq:tildekappa}\ee
We take $T=10$ and present results for $\tilde{\Theta}$ and  $\tilde{\kappa}$ in a 1D chain in \figs{Fig:tildelabmda1}{Fig:tildekappa1}. In general, for $\Gamma=1$ and $\Omega=0.5$, we find that for $J\mathcal{Z}\gtrsim 5$ there is a $\Delta$ region (increasing in width with $J\mathcal{Z}$) where the MFQF approximation breaks in 1D as the correlations become too large (a border that we define to be $\tilde{\Theta}>1$, a clearly unphysical value). 

\begin{figure}
\includegraphics[width=\linewidth]{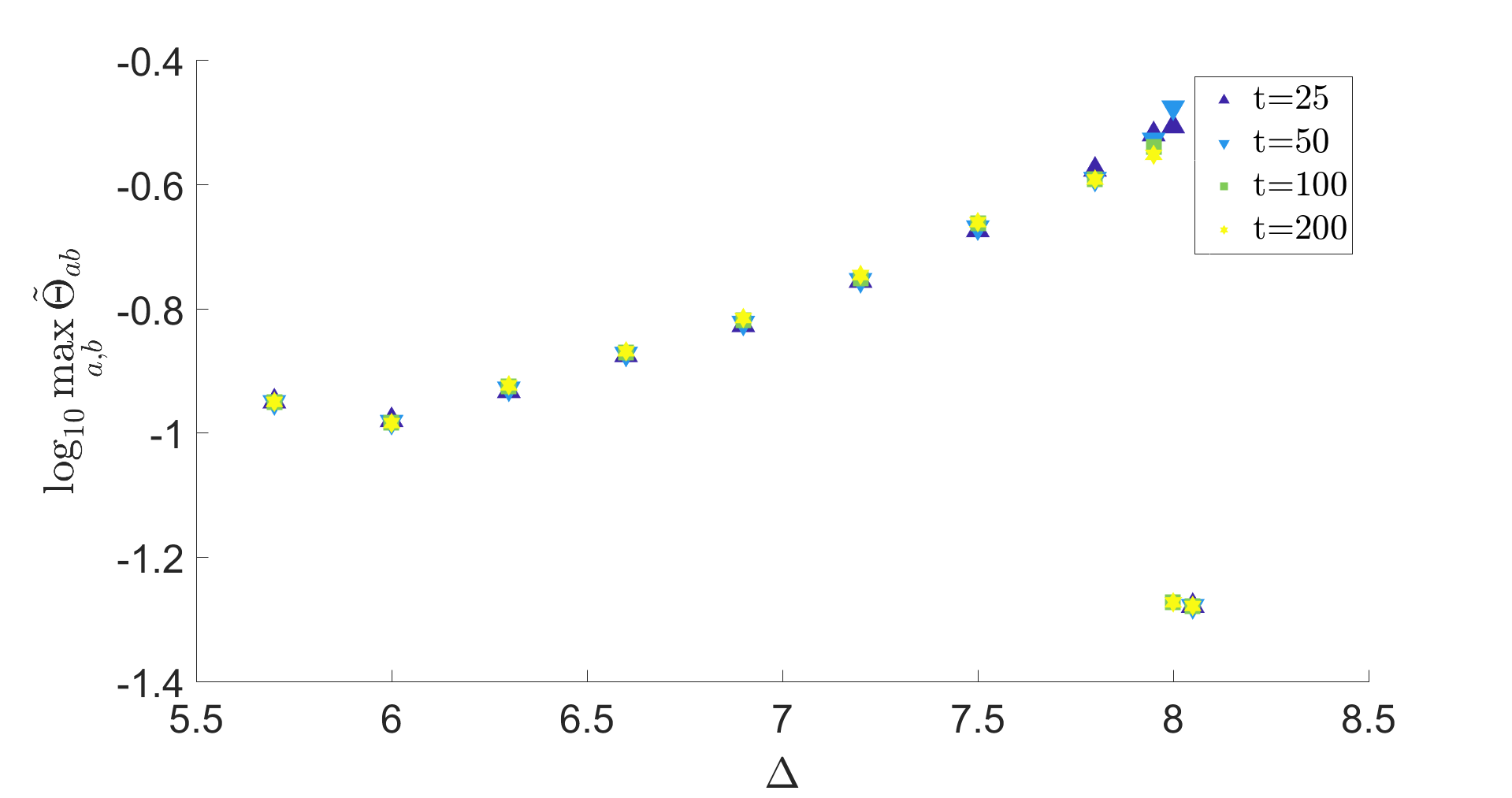}
\caption{The measure of convergence of the correlation functions used for MFQF [\eq{Eq:tildelambda}], for the intermediate branch with the parameters $\Gamma=1$, $\Omega=0.5$, $J\mathcal{Z}=10$ on a 3D lattice with $N=30^3$ sites.}
\label{Fig:tildelabmda3}
\end{figure}

\begin{figure}[t!]
\includegraphics[width=\linewidth]{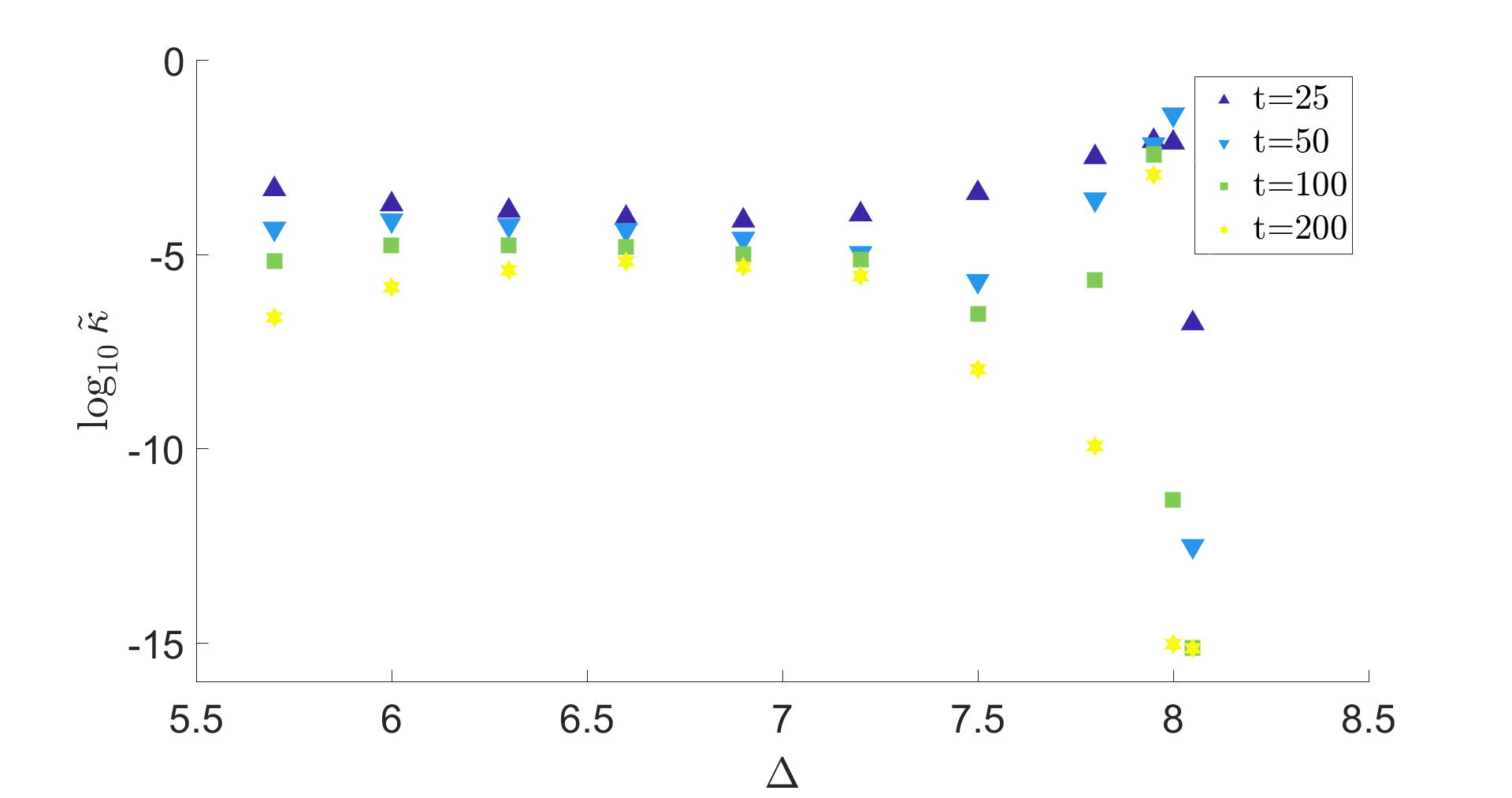}
\caption{The measure of convergence of the time dynamics used for MFQF [\eq{Eq:tildekappa}] results, for the intermediate branch with the parameters $\Gamma=1$, $\Omega=0.5$, $J\mathcal{Z}=10$ on a 3D lattice with $N=30^3$ sites.}
\label{Fig:tildekappa3}
\end{figure}

For 2D and higher dimensional lattices we did not find such cases, and the method has converged for the various parameter values that have been checked. In 2D we find that for the same parameters as in \fig{Fig:tildelabmda1}, $\tilde{\Theta}_{xx}\lesssim 0.6$ with other components reaching $\tilde{\Theta}_{ab}\sim 0.1$. The correlation length shown in \fig{Fig:Correlations2D}(a), which is at most 10-20 lattice sites up to the edge of the intermediate branch, implies that the simulations are well converged with the simulated lattice of $200^2$ sites. As the dimension is increased, $\tilde{\Theta}_{ab}$ decrease further, and also the typical correlation lengths $\lambda_{ab}$ for similar parameters decrease. In \figs{Fig:tildelabmda3}{Fig:tildekappa3} we show the measure of convergence for a lattice in 3D (simulated with $30^3$ sites), for the intermediate branch and the same parameters, with the maximal correlation $\tilde{\Theta}_{ab}\sim 0.3$. We also find that the correlation length is at most 5-6 lattice 
sites for most $\Delta$ values, until it starts to increase sharply, and in the range $7.2 \lesssim \Delta \lesssim 8$ the correlation length assumes a magnitude of the order of the lateral lattice size available in the simulations.

\section{Convergence checks of the MPO calculations}
\label{App:ConvMPO}

\begin{figure}[tb!]
\includegraphics[width=\linewidth]{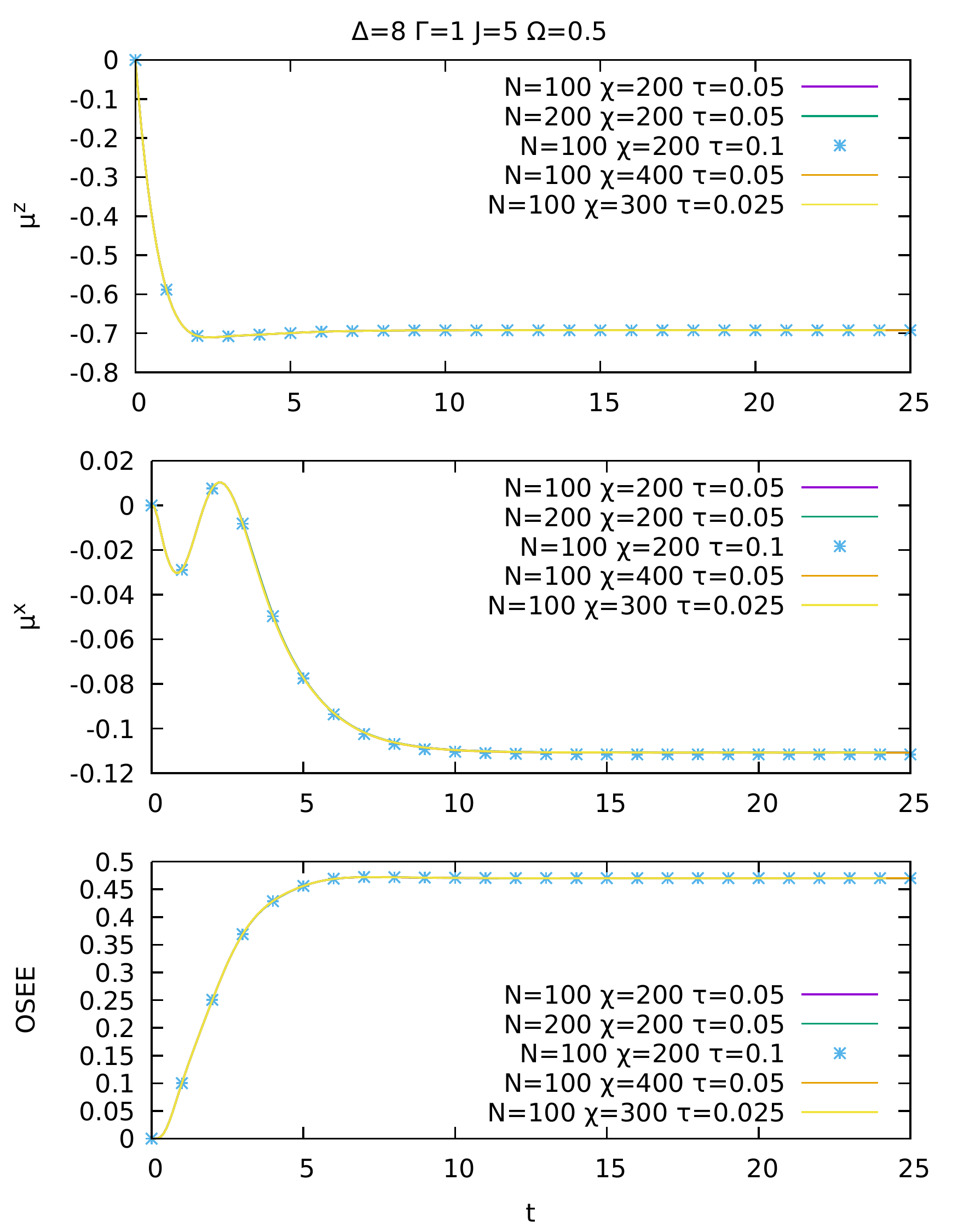}
\caption{Time evolution of the $z$ component of the magnetization, $x$ component, and OSEE
for various sets of MPO simulation parameters. It illustrates the good convergence of the results for  $\Delta=5$, $\Omega=0.5$, $\Gamma=1$ and $J=5$. At this scale, we observe that the results are practically unchanged if one varies the time step 
$\tau$ from 0.1 to 0.025, if one changes the maximum bond dimension $\chi$ from 200 to 400, or if one increases the system size $N$ from 100 to 200.
}
\label{Fig:ConvMPO2}
\end{figure}

\begin{figure}[tb!]
\includegraphics[width=\linewidth]{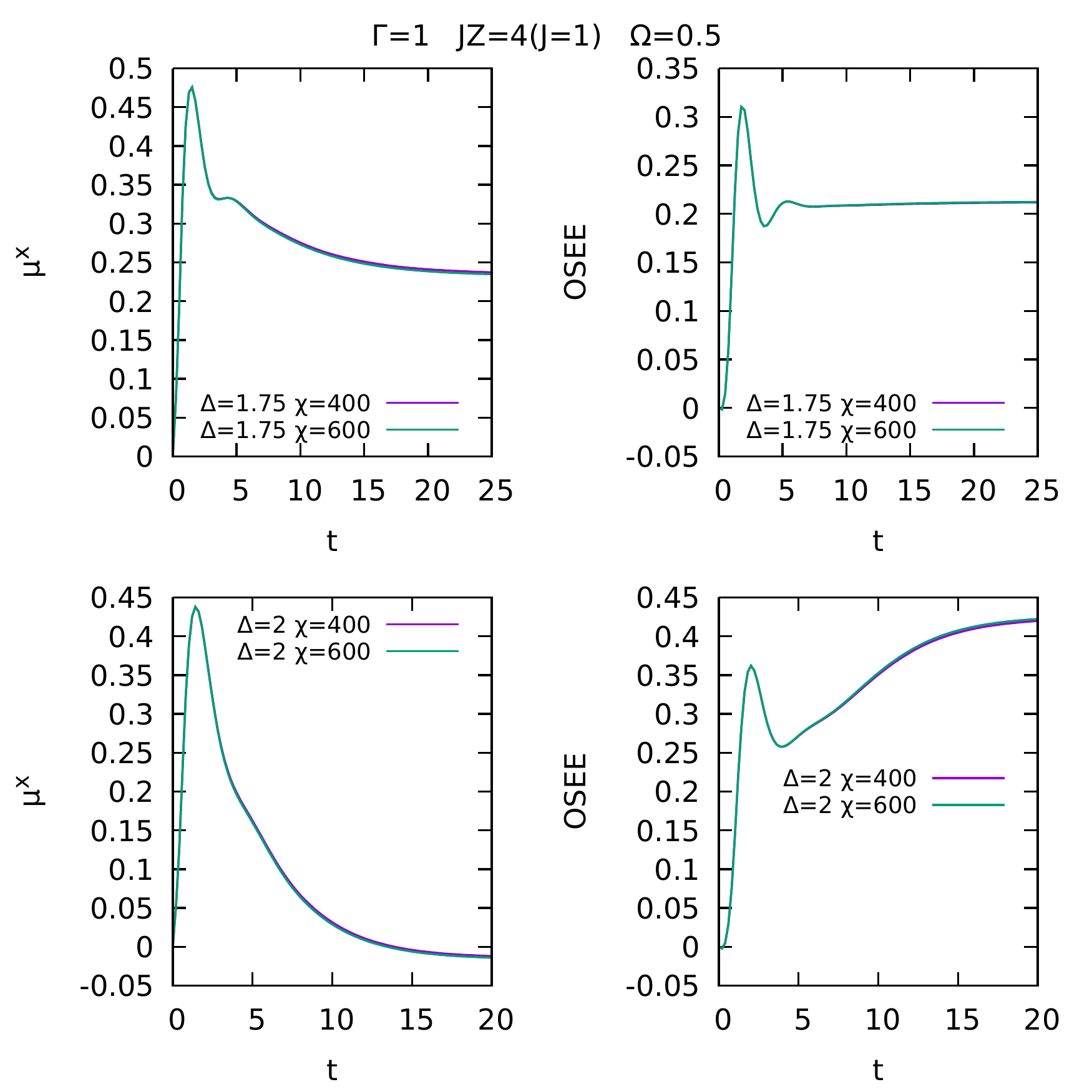}
\caption{Time evolution of the $z$ component of the magnetization and OSEE
for two values of the maximal  bond dimension $\chi$, in a $4\times 12$ cylinder.
$\Delta=1.75$ in the top panels, and $\Delta=2$ in the bottom ones.
Other parameters: $\Gamma=1$, $J=2.5$.
The curves associated to maximal bond dimensions $\chi=400$ and 600 are almost on top of each other at the scale of these plots.
}
\label{Fig:ConvMPO3}
\end{figure}

\begin{figure}[tb!]
\includegraphics[width=\linewidth]{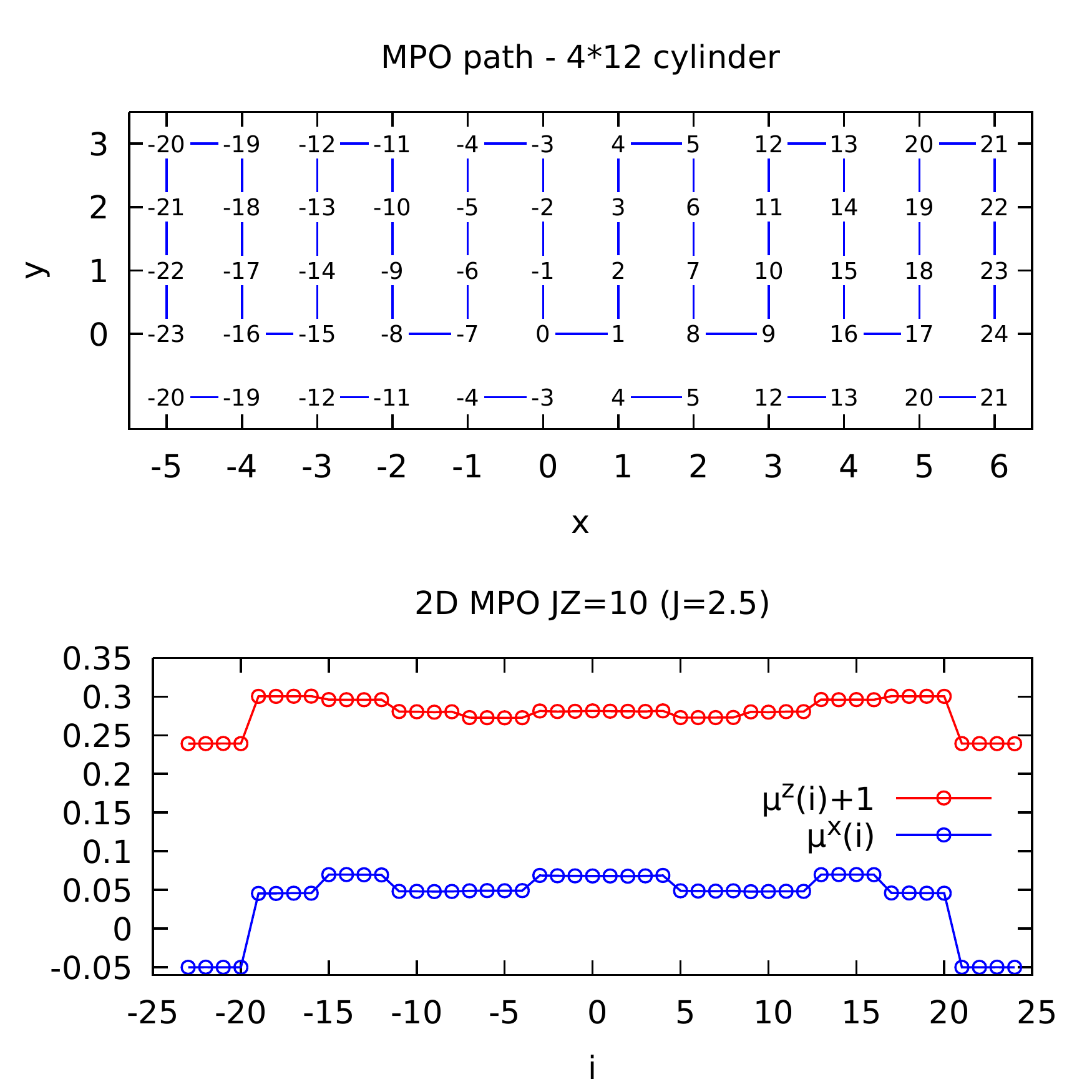}
\caption{Top: one-dimensional MPO path used to simulate the system on cylinder (periodic boundary condition in the $y$ direction, and
open ones in the $x$ direction). Bottom: local mean magnetizations $\mu^x(i)$ and $\mu^z(i)$ (shifted by 1 for clarity) in the steady state, as a function of the site index along the 1D path. The plateaus of width 4 reflect the fact that the magnetization is translation invariant in the $y$ direction, as it should be. To pass such a (sanity) check, the bond dimension should be large enough, since the translation invariance in the $y$ direction is explicitly broken by the 1D path.  Parameters: $J\mathcal{Z}=10$ ($J=2.5$), $\Delta=6$, $\Gamma=1$ and $\Omega=0.5$. Simulation parameters: maximum bond dimension $\chi=400$,
trotter step $\tau=0.05$ and total time evolution $t=20$.
}
\label{Fig:eta_mpo}
\end{figure}

Our MPO implementation is based on the iTensor library \cite{itensor}, and encodes the Liouvillian super operator as a super-MPO (acting on the state which is an MPO). We evolve $\rho$ in real time using a Trotter scheme of order 4 \cite{zaletel_time-evolving_2015,bidzhiev_out--equilibrium_2017}, with an error for each step which scale as  $\mathcal{O}(\tau^5)$.
In the present study we typically used $\tau=0.1$ or $0.05$ (depending on the magnitude of the model parameters).
Another crucial  parameter is the maximum [bond] dimension $\chi$ used to truncate the Schmidt spectra after each singular value decomposition. 
The errors that are introduced can be estimated by checking how the relevant observables change when varying the parameters above.
The results for 1D chains are summarized in Fig.~\ref{Fig:ConvMPO2}. The effect of the finite bond dimension is illustrated in \fig{Fig:ConvMPO3},
where a simulation with $\chi=400$ is compared to one with $\chi=600$.

\section{Transformation to the rotating frame}
\label{App:Rotating}

We consider a two-level system, represented by spin-1/2 operators. The energy difference between the two $\sigma^z$ eigenstates (with $\sigma^z$ eigenvalues 1 and -1) is modelled by the term 
\be H_0 = \omega_c \sigma^z / 2.\ee
 The driving term in the lab frame is a classical external field which couples to the spin, of amplitude $2\Omega$ and rotating at the angular frequency $\omega$. It corresponds to the following term: 
\be V(t) = 2\Omega \cos(\omega t)\sigma^x.\ee
This could, for instance, describe the rotating electric field of a laser (or of a microwave) coupled to the two-level system \cite{krantz2019quantum}. The mapping from this time-dependent Hamiltonian $H_{\rm lab}=H_0+V(t)$ in the lab frame to a  time-independent Hamiltonian (in the rotating frame) amounts to using the unitary transformation 
\be U(t)=\exp\{i\omega t \sigma^z /2 \},\ee and to considering the transformed density matrix $\tilde\rho(t)=U(t)^\dag  \rho(t) U(t)$. The Hamiltonian in the rotating frame is given by
\be H = U(t)^\dag H_{\rm lab} U(t)+i \partial_t U(t)^\dag U(t),\ee
which, making the rotating wave approximation (neglecting terms rotating with frequency $2\omega$, justified for $\omega\gg \Omega$), results in the time-independent Hamiltonian
\be H = \Omega \sigma^x + \Delta \sigma^z/2, \qquad \Delta \equiv \omega_c-\omega.\ee
As for the jump terms, a simple calculation shows that they are not affected by the unitary transformation. This is because a unitary rotation of angle $\theta$ about the $z$ axis transforms $\sigma^+$ into $\sigma^+ \exp(i\theta)$ and $\sigma^-$ into $\sigma^- \exp(-i\theta)$. Since a $\sigma^+$ operator is always accompanied with a $\sigma^-$ operator both in the coherent interaction terms and in the Lindblad dissipator terms, none are modified when going to the rotating frame.

\bibliography{etatheory}

\end{document}